\documentclass[journal]{IEEEtran}

\ifCLASSINFOpdf
\usepackage[pdftex]{graphicx}
\else
\usepackage[dvips]{graphicx}
\fi
\usepackage{amsmath}
\usepackage{mathrsfs}
\usepackage{amssymb} 
\usepackage{amsfonts}
\usepackage{multicol}
\usepackage{enumerate}
\usepackage{color}
\usepackage{multirow}
\usepackage{siunitx}
\usepackage{amssymb}

\begin{document}

\title{WFRFT-aided Power-efficient Multi-beam Directional Modulation Schemes Based on Frequency Diverse Array}

\author{Qian~Cheng,
        Vincent~Fusco,~\IEEEmembership{Fellow, IEEE},
        Jiang~Zhu,~\IEEEmembership{Member, IEEE},
        Shilian~Wang,~\IEEEmembership{Member, IEEE},
        and Fanggang~Wang,~\IEEEmembership{Senior Member, IEEE}
\thanks{\copyright  2019 IEEE.  Personal use of this material is permitted.  Permission from IEEE must be obtained for all other uses, in any current or future media, including reprinting/republishing this material for advertising or promotional purposes, creating new collective works, for resale or redistribution to servers or lists, or reuse of any copyrighted component of this work in other works.}
\thanks{The work of Q. Cheng was supported by a scholarship (No. 201803170247) from China Scholarship Council (CSC). The work of F. Wang was supported in part by the Beijing Natural Haidian Joint Fund under Grant L172020, in part by the National Natural Science Foundation under Grant 61571034 and Grant U1834210, in part by the Beijing Natural Science Foundation under Grant 4182051, in part by the State Key Laboratory of Rail Traffic Control and Safety under Grant RCS2019ZT011, and in part by the Major Projects of Beijing Municipal Science and Technology Commission under Grant Z181100003218010. \textit{(Corresponding authors: Shilian Wang; Vincent Fusco.)} }
\thanks{Q. Cheng is with the College of Electronic Science, National University of Defense Technology, Changsha 410073, China, and also with the Institute of Electronics, Communications and Information Technology (ECIT), Queen's University Belfast, Belfast BT3 9DT, U.K. (email: chengqian14a@nudt.edu.cn).}
\thanks{V. Fusco is with the Institute of Electronics, Communications and Information Technology (ECIT), Queen's University Belfast, Belfast BT3 9DT, U.K. (e-mail: v.fusco@ecit.qub.ac.uk).}
\thanks{J. Zhu and S. Wang are with the College of Electronic Science, National University of Defense Technology, Changsha 410073, China (e-mail: jiangzhu@nudt.edu.cn, wangsl@nudt.edu.cn).}
\thanks{F. Wang is with the State Key Laboratory of Rail Traffic Control and Safety, Beijing Jiaotong University, Beijing 100044, China (e-mail: wangfg@bjtu.edu.cn).}
\thanks{Digital Object Identifier 10.1109/TWC.2019.2934462}
}

\maketitle

\begin{abstract}
The artificial noise (AN) aided multi-beam directional modulation (DM) technology is capable of wireless physical layer secure (PLS) transmissions for multiple desired receivers in free space. The application of AN, however, makes it less power-efficient for such a DM system. To address this problem, the weighted fractional Fourier transform (WFRFT) technology is employed in this paper to achieve power-efficient multi-beam DM transmissions. Specifically, a power-efficient multi-beam WFRFT-DM scheme with cooperative receivers and a power-efficient multi-beam WFRFT-DM scheme with independent receivers are proposed based on frequency diverse array (FDA), respectively. The bit error rate (BER), secrecy rate, and robustness of the proposed multi-beam WFRFT-DM schemes are analyzed. Simulations demonstrate that 1) the proposed multi-beam WFRFT-DM schemes are more power-efficient than the conventional multi-beam AN-DM scheme; 2) the transmission security can also be guaranteed even if the eavesdroppers are located close to or the same as the desired receivers; and 3) the proposed multi-beam WFRFT-DM schemes are capable of independent transmissions for different desired receivers with different modulations.
\end{abstract}


\IEEEpeerreviewmaketitle

\section{Introduction}

\IEEEPARstart{A}{s} a promising wireless physical layer secure (PLS) transmission strategy, the directional modulation (DM) technology has developed rapidly over decades. The DM technology can produce a standard symbol constellation for the desired receiver along a specific direction in free space, while simultaneously distorting the received signal of the eavesdropper \cite{Ding_Review_DM}. 

Generally, there are two main methods for DM synthesis, i.e., the optimization method at the radio frequency (RF) frontend \cite{Daly_PA_DM1}-\cite{Cheng_Time_FDA_DM} and the artificial noise (AN) aided method in the baseband \cite{Ding_Vector_DM}-\cite{Ji_Fading_FDA_DM}. For each method, the synthesis for DM transmission can be achieved using phased array (PA) or frequency diverse array (FDA). The difference between these two kinds of arrays is that the PA-based DM schemes can only achieve angle-dependent wireless PLS transmissions while the FDA-based DM schemes can realize wireless PLS transmissions in both angle and range dimensions.

Specifically, the optimization approach of DM technology at the RF frontend was first proposed in \cite{Daly_PA_DM1}\cite{Daly_PA_DM2}, where a group of phase shifters were optimized to produce the standard baseband symbols along a given direction. A bit error rate (BER) driven DM synthesis method was presented in \cite{Ding_BER_DM} that the standard constellation mappings in in-phase and quadrature (IQ) space can be exactly achieved along a specific direction. A switched antenna array \cite{Hong_RF}, a dual-beam antenna array excited by the IQ baseband signals \cite{Hong_Dual_Beam}, and a sparse linear antenna array \cite{Hong_Convex_Opt} were proposed, respectively, to perform wireless DM transmissions at the RF frontend. In addition to the PA-based DM schemes, the optimization method at the RF frontend was also applied into the FDA  to achieve both angle-range dependent DM transmissions \cite{Wang_FDA_DM}-\cite{Cheng_Time_FDA_DM}.

The optimization algorithm at the RF frontend, however, is usually pretty complicated, which leads to the emerging of the AN-based synthesis method in the baseband. An orthogonal AN-based approach was proposed in \cite{Ding_Vector_DM} to perform DM synthesis in the baseband. The authors in \cite{Hu_Robust_DM} further put forward a robust DM synthesis method with the aid of AN. The AN-based synthesis method  was also combined with the FDA to implement both angle-range dependent wireless PLS transmissions \cite{Hu_Random_FDA_DM}\cite{Ji_Fading_FDA_DM} in the baseband. { Another AN-aided DM based on FDA was proposed in \cite{Qiu_AN_FDA_DM}, which addressed the PLS problem for close-located proximal legitimate user and eavesdroppers. }

Apart from the above-mentioned single-beam DM schemes with only one desired direction/receiver \cite{Daly_PA_DM1}-\cite{Qiu_AN_FDA_DM}, { the multi-beam DM synthesis approaches with multiple desired directions/receivers were also investigated intensively including the lin-of-sight (LoS) models  \cite{Ding_Orthogonal_MB}-\cite{Qiu_MB_FDA_DM} and multi-path models \cite{Ding_RDA_DM}-\cite{Hafez_SM_DM}.} Specifically, the orthogonal AN-based synthesis method was proposed in  \cite{Ding_Orthogonal_MB}, which is capable of concurrently projecting independent data streams into different specified spatial directions while simultaneously distorting signal constellations in all other directions.  { The authors in \cite{Xie_AN_MB} further put forward the zero-forcing (ZF) based approach for multi-beam DM synthesis with the aid of AN. The AN was also employed in \cite{Shu_AN_MB} and \cite{Christopher_AN_MB} to realize multi-beam DM transmissions using precoding and iterative convex optimization, respectively. The work in \cite{Shu_Robust_MB1}\cite{Shu_Robust_MB2} presented a robust synthesis method using optimization algorithms for multi-beam DM transmissions with imperfect estimations of directions. In addition to multi-direction DM transmissions, point-to-multipoint DM transmissions were studied in \cite{Cheng_SVD_DM}-\cite{Qiu_MB_FDA_DM}  using FDA.  The retrodirective phased array was studied in \cite{Ding_RDA_DM} to realize AN-aided secure DM transmission in multi-path environments. The multi-path nature of the wireless channel was also exploited in \cite{Hafez_Multipath_DM2}-\cite{Hafez_SM_DM} to provide location-specific secure communication to legitimate users. }

{ Nearly all of these multi-beam DM synthesis approaches in LoS channels, however, require AN in the transmitting signals. The inserted AN inevitably consumes a proportion of the transmitting power and consequently lowers the power efficiency of the multi-beam DM systems. This paper aims to address the problem of low power efficiency of the AN-based multi-beam DM systems in LoS channels by applying the weighted fractional Fourier transform (WFRFT) technology into multi-beam DM synthesis. }

The WFRFT technology, as a special variation of Fourier transform \cite{Mei_Research_on_4WFRFT}, has been widely used for wireless PLS transmissions \cite{Fang_Guaranteeing_PLS_WFRFT}-\cite{Wang_NOMA_WFRFT}. Specifically, a WFRFT-aided cooperative overlay system was proposed in \cite{Fang_Guaranteeing_PLS_WFRFT} to guarantee wireless PLS transmission. The WFRFT technology was combined with parallel combinatory spreading technology for secret communications in \cite{Fang_Secret_Commu_WFRFT},  where the transmission data was divided into two groups to generate pseudonoise sequence and encrypt the information data, respectively. In \cite{Luo_Polarization_WFRFT1}\cite{Luo_Polarization_WFRFT2}, the WFRFT was amalgamated with polarization modulation for secure dual-polarized satellite communications. The authors in \cite{Fang_PLS_Cooperation_WFRFT1}\cite{Fang_PLS_Cooperation_WFRFT2} modeled the WFRFT-based wireless PLS cooperation transmission problem as a coalitional game with non-transferable utility. The WFRFT technology with multiple parameters \cite{Ding_Alterable_WFRFT}\cite{Liang_Multi_Para_WFRFT} was applied into secure spatial modulation system in \cite{Cheng_SM_WFRFT}. Moreover,  the WFRFT technology was also applied into hybrid carrier system in \cite{Sha_Hybrid_CDMA_WFRFT}-\cite{Wang_NOMA_WFRFT} for wireless PLS transmissions. 

To the best of our knowledge, no work regarding the amalgamation of WFRFT and DM technologies has been investigated in the present literature for the purpose of improving the power efficiency of the AN-based multi-beam DM systems. Overall, the main contributions of our work are as follows: 

\begin{enumerate}
\item The WFRFTT technology is applied into DM systems to enhance the power efficiency of multi-beam DM systems for the first time, which outperforms the conventional multi-beam AN-DM scheme.

\item Two power-efficient multi-beam WFRFT-DM schemes with cooperative/independent receivers are proposed, respectively. For each scheme, the architectures of the transmitter and the legitimate desired receivers are elaborately designed, respectively.

\item {The BER, secrecy rate, and robustness of the proposed multi-beam WFRFT-DM schemes are deduced and verified with simulations. The comparisons among different multi-beam DM schemes are also provided. Some practical issues for the proposed multi-beam WFRFT-DM schemes are presented as well.}

\end{enumerate}

The rest of this paper is organized as follows. The principle of WFRFT is briefly introduced in Section II. Section III and Section IV illustrate the principles of the proposed power-efficient multi-beam WFRFT-DM schemes with cooperative and independent receivers, respectively. Section V deduces the BER, secrecy rate, and robustness of the proposed WFRFT-DM schemes; {and further compares different multi-beam DM schemes.} Finally, Section VI makes a conclusion for this paper and points out the future work.

\emph{Notations}: Operators ${( \cdot )^{\rm{*}}}$, ${( \cdot )^{\rm{T}}}$, ${( \cdot )^{\rm{H}}}$, ${( \cdot )^{-1}}$, and ${( \cdot )^{\dagger}}$ represent the complex conjugation, transpose, Hermitian transpose, inverse, and Moore-Penrose inverse operations of a matrix, respectively. Operators $\mathscr{D}( \cdot )$ and $\mathscr{F}( \cdot )$ stand for the normalized discrete Fourier transform (DFT) and the weighted fractional Fourier transform (WFRFT) of a sequence, respectively. The sets of integer, real, and complex numbers are denoted by $\mathbb{Z}$, $\mathbb{R}$, and $\mathbb{C}$, respectively. { The notations $\mathbb{E}(\cdot)$,  ${\rm tr}(\cdot)$,  $\max\{\cdot\}$, and $|\cdot|$ refer to the expectation of a random variable, the trace of a matrix,  the maximum value of a set of real numbers, and the modulus of a complex number, respectively. } $\mathbf{I}_{K}$ indicates the identity matrix with size $K\times K$, while ${\cal CN}(0,{\sigma ^2})$ refers to the complex Gaussian distribution with zero mean and variance ${\sigma ^2}$.

\section{Weighted Fractional Fourier Transform}

The weighted fractional Fourier transform is one kind of variations of Fourier transform. Particularly, the 4-order WFRFT is the weighted sum of the 1 to 4 times of normalized discrete Fourier transform (DFT) of a sequence \cite{Mei_Research_on_4WFRFT}, which has been widely used for wireless PLS transmissions.

The normalized DFT of an arbitrary ${J}$-length complex sequence, ${\bf{s}} = [s_0~s_1~\cdots~s_{{J}-1}]^{\rm{T}}\in {\mathbb{C}}^{{J}\times 1}$, is denoted by  $\mathop{\bf{s}}\limits^{.} = {\mathscr{D}}({\bf{s}}) = [{\mathop{s}\limits^{.}}_0~{\mathop{s}\limits^{.}}_1~\cdots~{\mathop{s}\limits^{.}}_{{J}-1}]^{\rm{T}}$ in this paper. The pair of normalized DFT and inverse DFT can be calculated by
\begin{equation}
\label{eq_DFT}
\left\{
\begin{array}{l}
{\mathop {s}\limits^{.}}_{k} = \frac{1}{\sqrt{{J}}}\sum\limits_{n=0}^{{J}-1} {s_{n}} \exp\left(-j\frac{2\pi}{J}kn\right)\\
{\mathop {s}_{n} }= \frac{1}{\sqrt{{J}}}\sum\limits_{k=0}^{{J}-1} {{\mathop {s}\limits^{.}}_{k}} \exp\left(j\frac{2\pi}{J}kn\right)
\end{array}
\right.
\end{equation} 

Using the normalized DFT in (\ref{eq_DFT}), the 4-order WFRFT can be defined as
\begin{equation}
\label{eq_WFRFT}
{\mathscr{F}^{\alpha }}({\bf{s}}) = {\omega _0}{\bf{s}} + {\omega _1}{\mathop{\bf{s}}\limits^{.}} + {\omega _2}{\mathop{\bf{s}}\limits^{..}} + {\omega _3}{\mathop{\bf{s}}\limits^{...}}
\end{equation} 
where $\alpha \in \mathbb{R}$ is a real WFRFT parameter; ${\mathop{\bf{s}}\limits^{.}}=\mathscr{D}({\bf{s}})$, ${\mathop{\bf{s}}\limits^{..}}=\mathscr{D}({\mathop{\bf{s}}\limits^{.}})$, ${\mathop{\bf{s}}\limits^{...}}=\mathscr{D}({\mathop{\bf{s}}\limits^{..}})$, and ${\bf{s}}=\mathscr{D}({\mathop{\bf{s}}\limits^{...}})$; $\omega_{i}$ ($i=0,1,2,3$) are the weight coefficients. For multi-parameter 4-order WFRFT, the $i$-th coefficient $\omega_{i}$ is defined as 
\begin{equation}
\label{eq_omega}
\omega_{i}=\frac{1}{4}\sum\limits_{k=0}^{3}\exp\left\{-j\frac{2\pi}{4}\left[(4m_k+1)(4n_k+k)\alpha-ki\right]\right\}
\end{equation}
where $M_V= [m_0~m_1~m_2~m_3]^{\rm{T}} \in \mathbb{Z}^{4\times 1}$ and $N_V= [n_0~n_1~n_2~n_3]^{\rm{T}} \in \mathbb{Z}^{4\times 1}$ are integer vectors. From (\ref{eq_WFRFT}) and (\ref{eq_omega}), the generalized multi-parameter 4-order WFRFT of a sequence is determined by nine independent parameters, i.e., $\alpha$, $M_V= [m_0~m_1~m_2~m_3]^{\rm{T}}$, and $N_V= [n_0~n_1~n_2~n_3]^{\rm{T}}$. These multiple independent parameters provide an opportunity to apply the 4-order WFRFT into  wireless PLS communications \cite{Ding_Alterable_WFRFT}\cite{Cheng_SM_WFRFT}.
Particularly, if we set $M_V=N_V= [0~0~0~0]^{\rm{T}}$, the weight coefficients will evolve into the single-parameter case \cite{Mei_Research_on_4WFRFT}, i.e.,
\begin{equation}
\label{eq_omega_single}
\omega_{i}=
\cos\left[\frac{(\alpha-i)\pi}{4}\right] 
\cos\left[\frac{2(\alpha-i)\pi}{4}\right]
\cos\left[\frac{3(\alpha-i)\pi}{4}\right]
\end{equation}

Before elaborating the principles of the proposed WFRFT-aided power-efficient multi-beam DM schemes, we summarize the properties of the 4-order WFRFT as follows  \cite{Mei_Research_on_4WFRFT}\cite{Fang_PLS_Cooperation_WFRFT2}:

$\bullet$ Boundary property:
\begin{equation}
\label{eq_WFRFT_Boundary}
{\mathscr{F}^{0}}({\bf{s}}) ={\bf{s}},~{\mathscr{F}^{1}}({\bf{s}}) ={\mathscr{D}}({\bf{s}})
\end{equation} 

$\bullet$ Periodical property:

The parameter $\alpha$ of the 4-order WFRFT has a periodicity of 4, i.e., 
\begin{equation}
\label{eq_WFRFT_Periodical}
{\mathscr{F}^{\alpha}}({\bf{s}}) ={\mathscr{F}^{\alpha +4}}({\bf{s}})
\end{equation} 

\begin{figure}
\centering
\includegraphics[angle=0,width=0.27\textwidth]{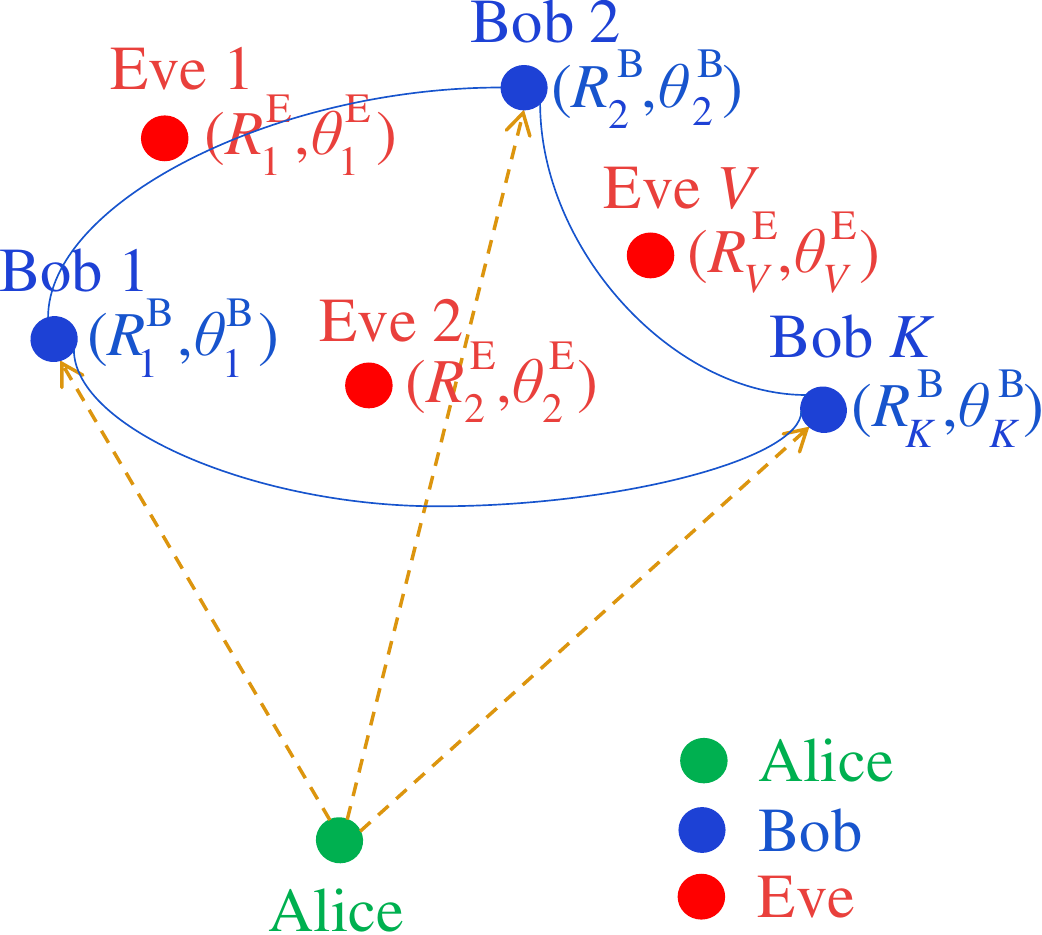}
\caption{The scenario of the proposed power-efficient multi-beam WFRFT-DM scheme with cooperative receivers.}
\end{figure}

$\bullet$ Additive property:

\begin{equation}
\label{eq_WFRFT_Addictive}
{\mathscr{F}^{\alpha + \beta}}(\bf{s}) = {\mathscr{F}^{\alpha }}[{\mathscr{F}^{\beta }} (\bf{s})] = {\mathscr{F}^{\beta }}[{\mathscr{F}^{ \alpha}} (\bf{s})]
\end{equation} 
where $\alpha,\beta\in \mathbb{R}$ are real numbers. For simplicity, we express ${\mathscr{F}^{\alpha }}[{\mathscr{F}^{\beta }} (\bf{s})]$ as ${\mathscr{F}^{\alpha }}{\mathscr{F}^{\beta }} (\bf{s})$ in the following analysis.

$\bullet$ Linear property:
\begin{equation}
\label{eq_WFRFT_Linear}
\left\{
\begin{array}{l}
{\mathscr{F}^{\alpha }}(c{\bf{s}}) = c{\mathscr{F}^{\alpha }} ({\bf{s}})\\
{\mathscr{F}^{\alpha }}({\bf{s}}+{\bf{u}}) = {\mathscr{F}^{\alpha }} ({\bf{s}})+{\mathscr{F}^{\alpha }} ({\bf{u}})
\end{array}
\right.
\end{equation} 
where ${\bf{s}},{\bf{u}}\in\mathbb{C}^{{J}\times1}$ are two complex sequences, and $c\in\mathbb{C}$  is a complex number.

From (\ref{eq_WFRFT_Boundary}) and (\ref{eq_WFRFT_Addictive}), the original sequence ${\bf{s}}$ can be recovered using the inverse WFRFT by simply substituting the WFRFT parameter $\alpha$ with $-\alpha$, i.e., ${\bf{s}}={\mathscr{F}^{-\alpha }}{\mathscr{F}^{\alpha }} (\bf{s})$.

\section{Power-efficient Multi-beam WFRFT-DM Scheme with Cooperative Receivers}

In this section, we consider a scenario where the legitimate transmitter (Alice) is trying to transmit confidential information to the $K$ legitimate receivers (Bobs) located at different places in free space, as shown in Fig. 1. {{These Bobs are assumed to be able to cooperate with each other by sharing their received signals, so that they can implement the inverse WFRFT with the same parameters together in order to obtain their own recovered signals.} Moreover, we assume there exist $V$ passive eavesdroppers (Eves).

{
\subsection{Fundamentals of Symmetrical Multi-Carrier FDA}
}

\begin{figure}
\centering
\includegraphics[angle=0,width=0.4\textwidth]{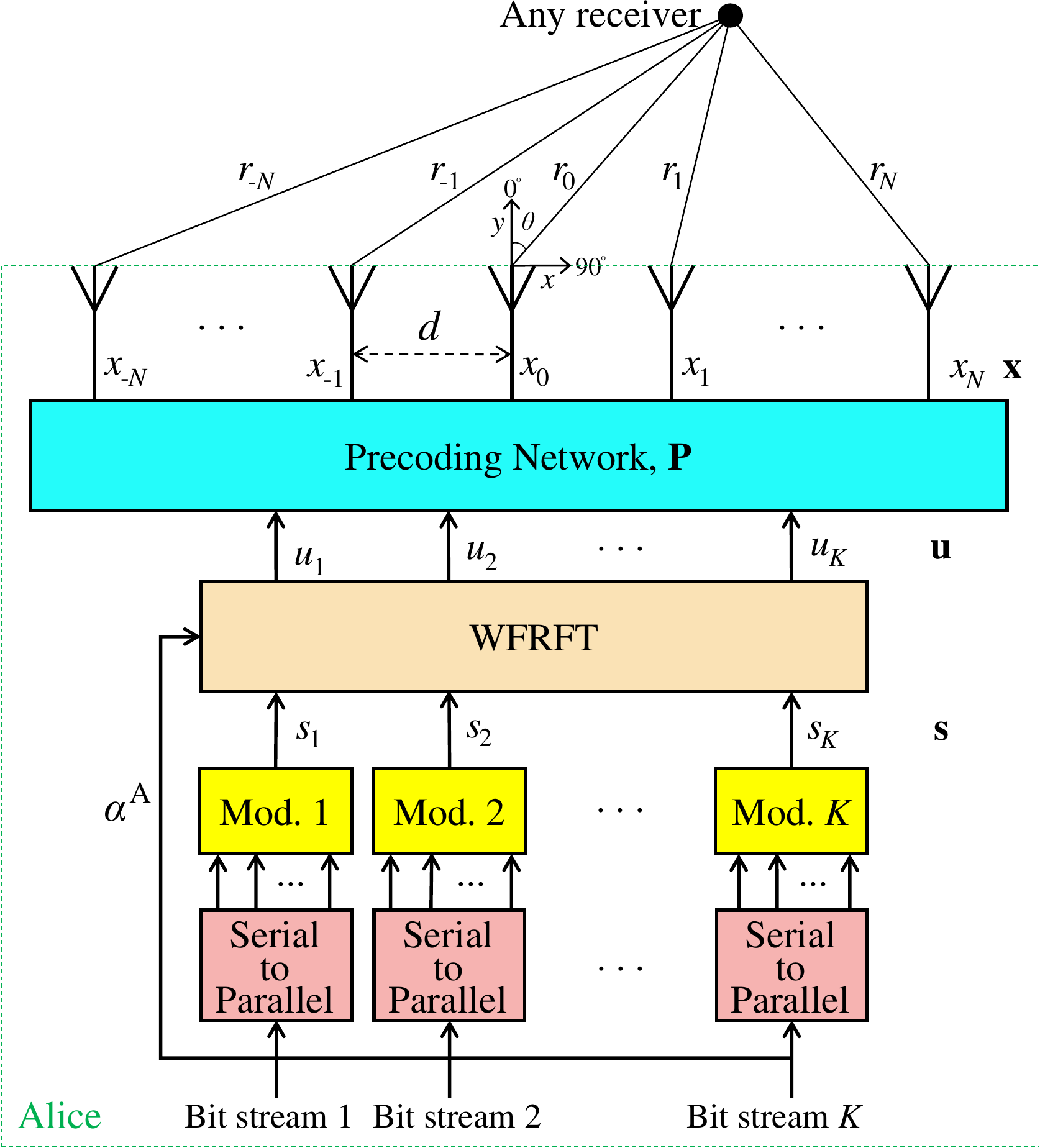}
\caption{The architecture of Alice for the proposed power-efficient multi-beam WFRFT-DM scheme with cooperative receivers.}
\end{figure}

The array architecture of the transmitter is shown in Fig. 2, which employs a ($2N+1$)-element symmetrical multi-carrier FDA with each antenna element emitting $L$ carriers \cite{Shao_MC_FDA}. The frequency increment between the central frequency $f_0$ and the $l$-th carrier ($l=0,1,\cdots,L-1$) of the $n$-th antenna element ($n=-N,\cdots,0,\cdots,N$) is designed as
\begin{equation}
\label{eq_Delta_f_n_l}
\Delta f_{n,l}=\Delta f\ln \left[(|n|+1)^{p}(l+1)\right]
\end{equation}    
where $\Delta f$ is a constant frequency increment, and $p$ is an extra parameter to control the frequency increments. The frequency increments should satisfy the following constraint, 
\begin{equation}
\label{eq_f_n_l_constraint}
|\Delta f_{n,l}|\ll f_{0}
\end{equation}

Therefore, the radiated frequency for the $l$-th carrier of the $n$-th antenna element is
\begin{equation}
\label{eq_f_n_l}
f_{n,l}=f_{0}+\Delta f_{n,l}
\end{equation}

{

In order to deduce the steering vector of the FDA, we consider that each element of the transmitter transmits sinusoidal signals with $L$ carriers. The signal transmitted by the the $n$-th element at time $t$ can  be expressed as
\begin{equation}
\label{eq3}
{x_n}(t) = \sum\limits_{l=0}^{L-1}e^{j2\pi {f_{n,l}}t}
\end{equation}

Let $r$ and $\theta$ represent the range and azimuth angle with respect to the central antenna element, respectively. For an arbitrary receiver located at $(r,\theta)$, 
the overall observed signal in the far field can be written as
\begin{subequations}
\label{y_r_theta1}
\begin{align}
y(r,\theta ) & = \sum\limits_{n = -N}^{N} {x_n}\left( {t - \frac{{{r_n}}}{c}} \right) \\
& =  \sum\limits_{n = -N}^{N} \sum\limits_{l=0}^{L-1} {\exp \left\{ {j{2\pi {f_{n,l}}\left( {t - \frac{{{r_n}}}{c}} \right)}} \right\}}
\end{align}
\end{subequations}  
where $c$ denotes light speed and $r_n$ refers to the path length from the $n$-th element to the observation point. With the far field approximation, $r_n$ can be expressed as
\begin{equation}
\label{far_field_approximation}
{r_n} \approx r - nd\sin \theta
\end{equation}
where $d$ represents the spacing between adjacent antenna elements, which is set as the half of central wavelength in this paper. Taking (\ref{far_field_approximation}) into (\ref{y_r_theta1}) yields (15) in the next page.

\newcounter{cnt1}
\setcounter{cnt1}{\value{equation}}
\setcounter{equation}{14}
\begin{figure*}[t]
\begin{subequations}
\label{y_r_theta2}
\begin{align}
y(r,\theta ) & \approx \sum\limits_{n = -N}^{N} \sum\limits_{l=0}^{L-1}{\exp \left\{ {j { 2\pi \left( {{f_0} + \Delta {f_{n,l}}} \right)\left( {t - \frac{{r - nd\sin \theta }}{c}} \right)} } \right\}} \\
& = \exp \left\{ {j2\pi {f_0}\left( {t - \frac{r}{c}} \right)} \right\} \sum\limits_{n = -N}^{N} \sum\limits_{l=0}^{L-1} {\exp \left\{ {j \left[{2\pi \Delta {f_{n,l}}\left( {t - \frac{r}{c}} \right) + \frac{{2\pi {f_0} nd\sin \theta }}{c} + \frac{{2\pi \Delta {f_{n,l}}nd\sin \theta }}{c}}\right] } \right\}}
\end{align}
\end{subequations}   
\end{figure*}
\setcounter{equation}{\value{cnt1}}
\setcounter{equation}{15} 

Moreover, the constraint in (\ref{eq_f_n_l_constraint}) implies  that (\ref{y_r_theta2}b) can be further approximated as
\begin{equation}
\label{y_r_theta3}
\begin{aligned}
&y(r,\theta ) \approx \exp \left\{ {j2\pi {f_0}\left( {t - \frac{r}{c}} \right)} \right\}  \\
&~~~\cdot\sum\limits_{n = -N}^{N} \sum\limits_{l=0}^{L-1} {\exp \left\{ {j \left[{2\pi \Delta {f_{n,l}}\left( {t - \frac{r}{c}} \right) + \frac{{2\pi {f_0} nd\sin \theta }}{c}} \right]} \right\}}
\end{aligned}
\end{equation}

\newcounter{cnt2}
\setcounter{cnt2}{\value{equation}}
\setcounter{equation}{16}
\begin{figure*}[t]
\begin{equation}
\label{eq_a_n_r_theta}
\begin{aligned}
{\mathbf{a}}_{n}(r,\theta)=\left[ 
e^{j\left(2\pi \Delta f_{n,0}\left (t-\frac{r}{c}\right)+\frac{ 2\pi f_{0}nd\sin \theta}{c}\right)}  
\cdots \right. e^{j\left(2\pi \Delta f_{n,l}\left (t-\frac{r}{c}\right)+\frac{2\pi f_{0}nd\sin \theta}{c}\right)} 
\cdots \left. e^{j\left(2\pi \Delta f_{n,L-1}\left (t-\frac{r}{c}\right)+\frac{2\pi f_{0}nd\sin \theta}{c}\right)} 
\right ]^{\rm{T}}
\end{aligned}
\end{equation}
\hrulefill
\end{figure*}
\setcounter{equation}{\value{cnt2}}
\setcounter{equation}{17}

In (\ref{y_r_theta3}), the terms inside the summation sign are decided by the geometry and the frequency-offset scheme of the system, therefore sub-steering vector caused by the $L$ carriers of the $n$-th antenna element can be written as (17) in the next page \cite{Shao_MC_FDA}\cite{Nusenu_Review_FDA}, which is an $L \times 1$   vector. As a result, the overall normalized steering vector of the symmetrical multi-carrier FDA can be calculated by 
\begin{equation}
\label{eq_h_r_theta}
\begin{aligned}
{\mathbf{h}}(r,\theta)=\frac {1}{\sqrt{(2N+1)L}} \left [{\mathbf{a}}_{-N}^{\rm{T}}(r,\theta) \cdots {\mathbf{a}}_{n}^{\rm{T}}(r,\theta) \cdots {\mathbf{a}}_{N}^{\rm{T}}(r,\theta) \right]^{\rm{T}}
\end{aligned}
\end{equation} 
which is a ($2N+1$)$L$$\times$$1$ vector.

}

\subsection{The Architecture of Alice with Cooperative Receivers}

The pre-transmitting baseband symbol $s_k$ for the $k$-th Bob ($k=1,2,\cdots,K$) is valued from the alphabet $\Omega_k$ with size $ |\Omega_k|=M_k$, i.e., $s_k \in \Omega_k$. The baseband symbols are assumed to be normalized, which means ${\mathbb{E}}[|s_k|^2]=1$. A symbol vector, $\mathbf{s}=\left[s_1~s_2~\cdots~s_K\right]^{\rm{T}}$, can be produced by combining the pre-transmitting baseband symbols for the $K$ Bobs. Then, the WFRFT with the parameter $\alpha^{\rm{A}}$ is operated to the symbol vector $\mathbf{s}$, which yields
\begin{equation}
\label{eq_u_WFRFT_s}
{\bf{u}} = {\mathscr{F}^{{\alpha ^{\rm{A}}}}}({\bf{s}}) = {\omega _0}{\bf{s}} + {\omega _1}{\mathop{\bf{s}}\limits^{.}} + {\omega _2}{\mathop{\bf{s}}\limits^{..}} + {\omega _3}{\mathop{\bf{s}}\limits^{...}}
\end{equation} 
where ${\mathop{\bf{s}}\limits^{.}}=\mathscr{D}({\bf{s}})$, ${\mathop{\bf{s}}\limits^{..}}=\mathscr{D}({\mathop{\bf{s}}\limits^{.}})$, ${\mathop{\bf{s}}\limits^{...}}=\mathscr{D}({\mathop{\bf{s}}\limits^{..}})$, and ${\bf{s}}=\mathscr{D}({\mathop{\bf{s}}\limits^{...}})$. According to the boundary and additive  properties of WFRFT in (\ref{eq_WFRFT_Boundary}) and (\ref{eq_WFRFT_Addictive}), the symbol vector $\mathbf{s}$ can be calculated by ${\bf{s}} = {\mathscr{F}^{ - {\alpha ^{\rm{A}}}}}({\bf{u}})$.

Before transmitting, the symbol vector $\mathbf{u}$ should be processed with a precoding matrix to match the $2N+1$ antenna elements. We assume that the spatial locations of the $K$ cooperative Bobs are prior known for Alice, the combined set of which can be expressed as
\begin{equation}
\label{eq_R_Theta_B}
(\Upsilon^{\rm{B}},\Theta^{\rm{B}})=\left\{({R_1^{\rm{B}}},\theta_1^{\rm{B}}),({R_2^{\rm{B}}},\theta_2^{\rm{B}}),\cdots,({R_K^{\rm{B}}},\theta_K^{\rm{B}})\right\}
\end{equation}
where $({R_k^{\rm{B}}},\theta_k^{\rm{B}})$, $k=1,2,\cdots,K$, is the spatial position of the $k$-th Bob. Then the steering vectors of these Bobs can compose a steering matrix, i.e.,
\begin{equation}
\label{eq_H_R_Theta_B}
\mathbf{H}(\Upsilon^{\rm{B}},\Theta^{\rm{B}})=\left[\mathbf{h}({R_1^{\rm{B}}},\theta_1^{\rm{B}})~\mathbf{h}({R_2^{\rm{B}}},\theta_2^{\rm{B}})~\cdots~\mathbf{h}({R_K^{\rm{B}}},\theta_K^{\rm{B}})\right]
\end{equation} 

{
Let $\mathbf{P}$ denote the Moore-Penrose inversion of $\mathbf{H}^{\rm{H}}(\Upsilon^{\rm{B}},\Theta^{\rm{B}})$, that is,
\begin{subequations}
\label{eq_P}
\begin{align}
\mathbf{P} & =\left[\mathbf{H}^{\rm{H}}(\Upsilon^{\rm{B}},\Theta^{\rm{B}})\right]^{\dagger}\\
& = \mathbf{H}(\Upsilon^{\rm{B}},\Theta^{\rm{B}})\left[\mathbf{H}^{\rm{H}}(\Upsilon^{\rm{B}},\Theta^{\rm{B}})\mathbf{H}(\Upsilon^{\rm{B}},\Theta^{\rm{B}})\right]^{-1}
\end{align}
\end{subequations} 
which is normalized to the conjugation of the steering matrix, i.e.,
\begin{equation}
\label{eq_P_H_relation}
{{\bf{H}}^{\rm{H}}}({\Upsilon ^{\rm{B}}},{\Theta ^{\rm{B}}}){\bf{P}} = {{\bf{I}}_K}
\end{equation} 

Then the normalized precoding matrix $\tilde{\mathbf{P}}$ at the transmitter can be designed as \cite{Xie_AN_MB}
\begin{equation}
\label{eq_P_normalized}
\tilde{\mathbf{P}}=\frac{1}{\sqrt{\rm{tr}(\mathbf{PP}^{\rm{H}})}}\cdot\mathbf{P}\triangleq\sqrt{\varepsilon}\mathbf{P}
\end{equation} 
where $\varepsilon\triangleq1/{\rm{tr}(\mathbf{PP}^{\rm{H}})}$ is a power normalization factor. After precoding, the radiating signal ${\bf{x}}=$ ${[{x_{ - N}}~ \cdots ~{x_0}~ \cdots ~{x_N}]^{\rm{T}}}$ for the $2N+1$ antenna elements can be obtained by
\begin{equation}
\label{eq_x}
{\bf{x}} = \sqrt {{\tilde{P}_s}} {\tilde{\mathbf{P}}\bf{u}} = \sqrt {\varepsilon{\tilde{P}_s}} {\bf{Pu}}\triangleq \sqrt {{{P}_s}} {\bf{Pu}}
\end{equation} 
where $\tilde{P}_s$ and $P_s\triangleq\varepsilon{\tilde{P}_s}$ refer to the signal power before and after precoding, respectively.
}

\subsection{The Architecture of Bob with Cooperative Receivers}

\begin{figure}
\centering
\includegraphics[angle=0,width=0.33\textwidth]{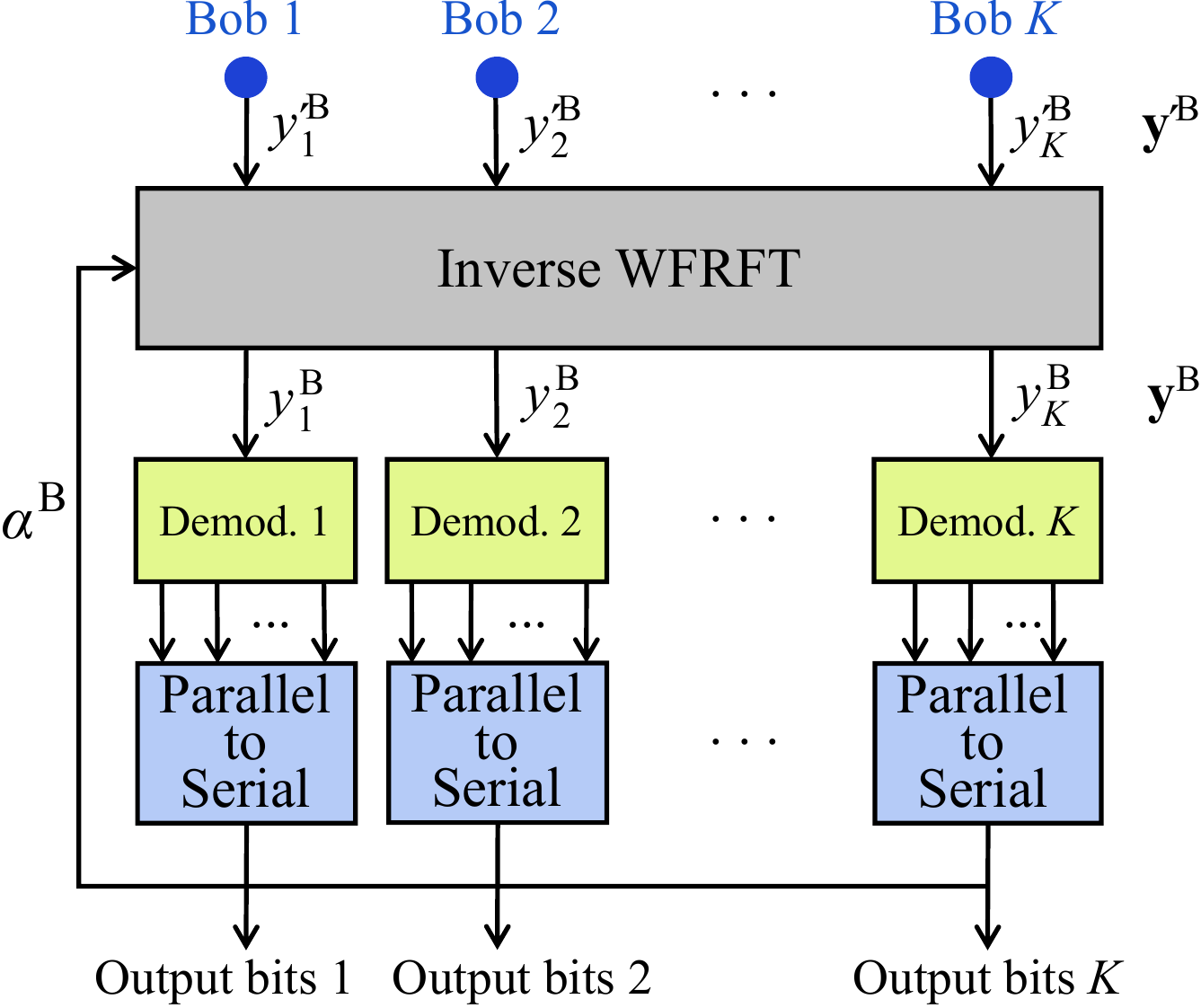}
\caption{The architecture of Bob for the proposed power-efficient multi-beam WFRFT-DM scheme with cooperative receivers.}
\end{figure}

The architecture of Bob for the proposed WFRFT-DM scheme with cooperative
receivers is shown in Fig. 3. The normalized LoS channel in free space is considered throughout this paper. After passing through the LoS channel, the received signal of the $k$-th Bob is obtained by
\begin{equation}
\label{eq_y_r_theta_k_B1}
{y'}_k^{\rm{B}} = y'({R_k^{\rm{B}}},\theta _k^{\rm{B}}) = {{\bf{h}}^{\rm{H}}}({R_k^{\rm{B}}},\theta _k^{\rm{B}}){\bf{x}} + {\xi'}_k^{\rm{B}}
\end{equation} 
where ${\xi'}_k^{\rm{B}} \sim {\cal CN}(0,{\sigma_{\xi^{\rm{B}}} ^2})$ is the complex additive white Gaussian noise (AWGN) with zero mean and variance ${\sigma_{\xi^{\rm{B}}} ^2}$. Since the $K$ Bobs are able to cooperate with each other, we can express their received signals as a vector, i.e.,
\begin{subequations}
\label{eq_Y_R_Theta_B1}
\begin{align}
{{{\bf{y'}}}^{\rm{B}}} &= {\bf{y}}'({\Upsilon ^{\rm{B}}},{\Theta ^{\rm{B}}}) = {\left[ {{y'}_1^{\rm{B}} ~~{y'}_2^{\rm{B}}~~ \cdots ~~{y'}_K^{\rm{B}}} \right]^{\rm{T}}}\\
&= {{\bf{H}}^{\rm{H}}}({\Upsilon ^{\rm{B}}},{\Theta ^{\rm{B}}}){\bf{x}} + {{{\boldsymbol{\xi}}'}^{\rm{B}}}
\end{align}
\end{subequations} 
where ${{{\boldsymbol{\xi}}'}^{\rm{B}}}= \left[{\xi'}_1^{\rm{B}}~{\xi'}_2^{\rm{B}}~\cdots~{\xi'}_K^{\rm{B}} \right]^{\rm{T}}  \sim {\cal CN}({\bf{0}}_{K\times 1},{\sigma_{\xi^{\rm{B}}} ^2}{\bf{I}}_K)$ is the AWGN vector.

We assume the WFRFT parameter of Alice is perfectly shared with Bobs for convenience. For these Bobs, their received signals are operated by the inverse WFRFT with the shared parameter, i.e.,
\begin{equation}
\label{eq_alpha_B}
{\alpha ^{\rm{B}}} =  - {\alpha ^{\rm{A}}}
\end{equation} 

Therefore, the received signals after inverse WFRFT can be expressed as
\begin{subequations}
\label{eq_Y_R_Theta_B2_1}
\begin{align}
{{\bf{y}}^{\rm{B}}} &= {\bf{y}}({\Upsilon ^{\rm{B}}},{\Theta ^{\rm{B}}}) = {\mathscr{F}^{{\alpha ^{\rm{B}}}}}\left( {{\bf{y'}}({\Upsilon ^{\rm{B}}},{\Theta ^{\rm{B}}})} \right)\\
&= {\mathscr{F}^{{\alpha ^{\rm{B}}}}}\left( {{{\bf{H}}^{\rm{H}}}({\Upsilon ^{\rm{B}}},{\Theta ^{\rm{B}}}){\bf{x}} + {{{{\boldsymbol{\xi}}'}}^{\rm{B}}}} \right)\\
 &= {\mathscr{F}^{{\alpha ^{\rm{B}}}}}\left( {\sqrt {{P_s}} {{\bf{H}}^{\rm{H}}}({\Upsilon ^{\rm{B}}},{\Theta ^{\rm{B}}}){\bf{Pu}}} \right) + {\mathscr{F}^{{\alpha ^{\rm{B}}}}}\left( {{{{{\boldsymbol{\xi}}'}}^{\rm{B}}}} \right)
\end{align}
\end{subequations}

Substituting (\ref{eq_u_WFRFT_s}), (\ref{eq_P_H_relation}) and  (\ref{eq_alpha_B}) into (\ref{eq_Y_R_Theta_B2_1}c), the received signals after inverse WFRFT can be further simplified as
\begin{subequations}
\label{eq_Y_R_Theta_B2_2}
\begin{align}
{{\bf{y}}^{\rm{B}}} &= {\mathscr{F}^{ {\alpha ^{\rm{B}}}}}\left(\sqrt {{P_s}} {\bf{I}}_{K} {\bf{u}} \right) + {\mathscr{F}^{{\alpha ^{\rm{B}}}}}\left( {{{{{\boldsymbol{\xi}}'}}^{\rm{B}}}} \right)\\
& = \sqrt {{P_s}} {\mathscr{F}^{ - {\alpha ^{\rm{A}}}}}{\mathscr{F}^{{\alpha ^{\rm{A}}}}}\left( {\bf{s}} \right) + {{{\boldsymbol{\xi}}}^{\rm{B}}}= \sqrt {{P_s}} {\bf{s}} + {{{\boldsymbol{\xi}}}^{\rm{B}}}
\end{align}
\end{subequations} 
where ${{{\boldsymbol{\xi}}}^{\rm{B}}}= \left[{\xi}_1^{\rm{B}}~{\xi}_2^{\rm{B}}~\cdots~{\xi}_K^{\rm{B}} \right]^{\rm{T}}$ is the noise vector after inverse WFRFT. Since the WFRFT operation does not change the distribution characteristics of AWGN \cite{Mei_Research_on_4WFRFT}, ${{{\boldsymbol{\xi}}}^{\rm{B}}} \sim {{\cal C}{\cal N}}({{\bf{0}}_{K \times 1}},\sigma _{{\xi}^{\rm{B}}}^2{{\bf{I}}_K})$ also holds.

Therefore, the received signal for the $k$-th Bob can be written as
\begin{equation}
\label{eq_y_r_theta_k_B2}
{y}_k^{\rm{B}} = y({R_k^{\rm{B}}},\theta _k^{\rm{B}}) = \sqrt {{P_s}} {{s}_{k}} + {\xi}_{k}^{\rm{B}}
\end{equation} 
which is simply the summation of the useful signal and AWGN. From (\ref{eq_y_r_theta_k_B2}), the $k$-th Bob can easily recover the confidential information transmitted from Alice.

\subsection{Eves' Received Signals with Cooperative Receivers}

\newcounter{cnt3}
\setcounter{cnt3}{\value{equation}}
\setcounter{equation}{31}
\begin{figure*}[t]
\begin{subequations}
\label{eq_y_r_theta_v_E}
\begin{align}
y({R_v^{\rm{E}}},\theta _v^{\rm{E}}) &= {{\bf{h}}^{\rm{H}}}({R_v^{\rm{E}}},\theta _v^{\rm{E}}){\bf{x}} + \xi_v^{\rm{E}} = \sqrt {{P_s}} {{\bf{h}}^{\rm{H}}}({R_v^{\rm{E}}},\theta _v^{\rm{E}}){\bf{P}}({\omega _0}{\bf{s}} + {\omega _1}{\mathop{\bf{s}}\limits^{.}} + {\omega _2}{\mathop{\bf{s}}\limits^{..}} + {\omega _3}{\mathop{\bf{s}}\limits^{...}}) + \xi_v^{\rm{E}}\\
 &= \underbrace {\sqrt {{P_s}} {\omega _0}{{\bf{h}}^{\rm{H}}}({R_v^{\rm{E}}},\theta _v^{\rm{E}}){\bf{Ps}}}_{{\rm{Distorted~Signal}}} + \underbrace {\sqrt {{P_s}} {{\bf{h}}^{\rm{H}}}({R_v^{\rm{E}}},\theta _v^{\rm{E}}){\bf{P}}({\omega _1}{\mathop{\bf{s}}\limits^{.}} + {\omega _2}{\mathop{\bf{s}}\limits^{..}} + {\omega _3}{\mathop{\bf{s}}\limits^{...}})}_{{\rm{Equivalent~AN}}} + \underbrace {\xi_v^{\rm{E}}}_{{\rm{AWGN}}}\\
 &= \underbrace {\sqrt {{P_s}} {\omega _0}{{\bf{h}}^{\rm{H}}}({R_v^{\rm{E}}},\theta _v^{\rm{E}}){\bf{Ps}}}_{{\rm{Distorted~Signal}}} + \underbrace {\sqrt {{P_s}} {{\bf{h}}^{\rm{H}}}({R_v^{\rm{E}}},\theta _v^{\rm{E}}){\bf{P}\boldsymbol{\eta }}}_{{\rm{Equivalent~AN}}} + \underbrace {\xi_v^{\rm{E}}}_{{\rm{AWGN}}}
\end{align}
\end{subequations} 
\hrulefill
\end{figure*}
\setcounter{equation}{\value{cnt3}}
\setcounter{equation}{32} 

But for the $v$-th Eve located at $({R_v^{\rm{E}}},\theta_{v}^{\rm{E}})$, $v=1,2,\cdots,V$, the received signal can be calculated by (\ref{eq_y_r_theta_v_E}) in the next page, where ${\xi}_v^{\rm{E}} \sim {\cal CN}(0,{\sigma_{\xi^{\rm{E}}} ^2})$ is the AWGN with zero mean and variance ${\sigma_{\xi^{\rm{E}}} ^2}$, and ${\boldsymbol{\eta}} = {\omega _1}{\mathop{\bf{s}}\limits^{.}} + {\omega _2}{\mathop{\bf{s}}\limits^{..}} + {\omega _3}{\mathop{\bf{s}}\limits^{...}}$ is the equivalent AN with variance $\sigma _\eta ^2 = 1-|\omega_0|^2$ \cite{Fang_PLS_Cooperation_WFRFT2}. In a worse case where the $V$ Eves are able to cooperate with each other, the received signal vector for the $V$ Eves can be written as 
\begin{subequations} 
\label{eq_Y_R_Theta_E}
\begin{align}
&{\bf{y}}^{\rm{E}}={\bf{y}}({\Upsilon ^{\rm{E}}},{\Theta ^{\rm{E}}}) = {{\bf{H}}^{\rm{H}}}({\Upsilon ^{\rm{E}}},{\Theta ^{\rm{E}}}){\bf{x}} + {\boldsymbol{\xi}^{\rm{E}}}\\
 &= \underbrace {\sqrt {{P_s}} {\omega _0}{{\bf{H}}^{\rm{H}}}({\Upsilon ^{\rm{E}}},{\Theta ^{\rm{E}}}){\bf{Ps}}}_{{\rm{Distorted~Signal}}} + \underbrace {\sqrt {{P_s}} {{\bf{H}}^{\rm{H}}}({\Upsilon ^{\rm{E}}},{\Theta ^{\rm{E}}}){\bf{P\boldsymbol{\eta} }}}_{{\rm{Equivalent~AN}}} + \underbrace {{\boldsymbol{\xi}^{\rm{E}}}}_{{\rm{AWGN}}}
\end{align}
\end{subequations} 
where ${{{\boldsymbol{\xi}}}^{\rm{E}}}= \left[{\xi}_1^{\rm{E}}~{\xi}_2^{\rm{E}}~\cdots~{\xi}_V^{\rm{E}} \right]^{\rm{T}}$ is the AWGN vector, and $\mathbf{H}(\Upsilon^{\rm{E}},\Theta^{\rm{E}})$ is the combined steering matrix of Eves, i.e.,
\begin{equation} 
\label{eq_H_R_Theta_E}
\mathbf{H}(\Upsilon^{\rm{E}},\Theta^{\rm{E}})=\left[\mathbf{h}({R_1^{\rm{E}}},\theta_1^{\rm{E}})~\mathbf{h}({R_2^{\rm{E}}},\theta_2^{\rm{E}})~\cdots~\mathbf{h}({R_V^{\rm{E}}},\theta_V^{\rm{E}})\right]
\end{equation} 

It is observed from (\ref{eq_y_r_theta_v_E}) and (\ref{eq_Y_R_Theta_E}) that the received signals of Eves are composed by three components. The first is the distorted signal, the second is the equivalent AN caused by WFRFT, and the third is the AWGN. Although there is no actual AN inserted in the transmitting signals, the application of WFRFT can produce equivalent AN for Eves, which means the proposed multi-beam WFRFT-DM scheme is more power-efficient than the conventional AN-DM scheme.

Owing to that the passive Eves are ignorant of the WFRFT parameters shared between Alice and Bobs, they cannot recover the confidential information. Moreover, even if the Eves' locations are close to or the same as Bobs', the transmission security can  be also guaranteed owing to the inherent security of WFRFT, which also outperforms the conventional multi-beam AN-DM scheme. For simplicity, we call the transmission security with close or the same locations of Eves and Bobs as \emph{neighbor security} in the following analysis.

\begin{figure}
\centering
\includegraphics[angle=0,width=0.27\textwidth]{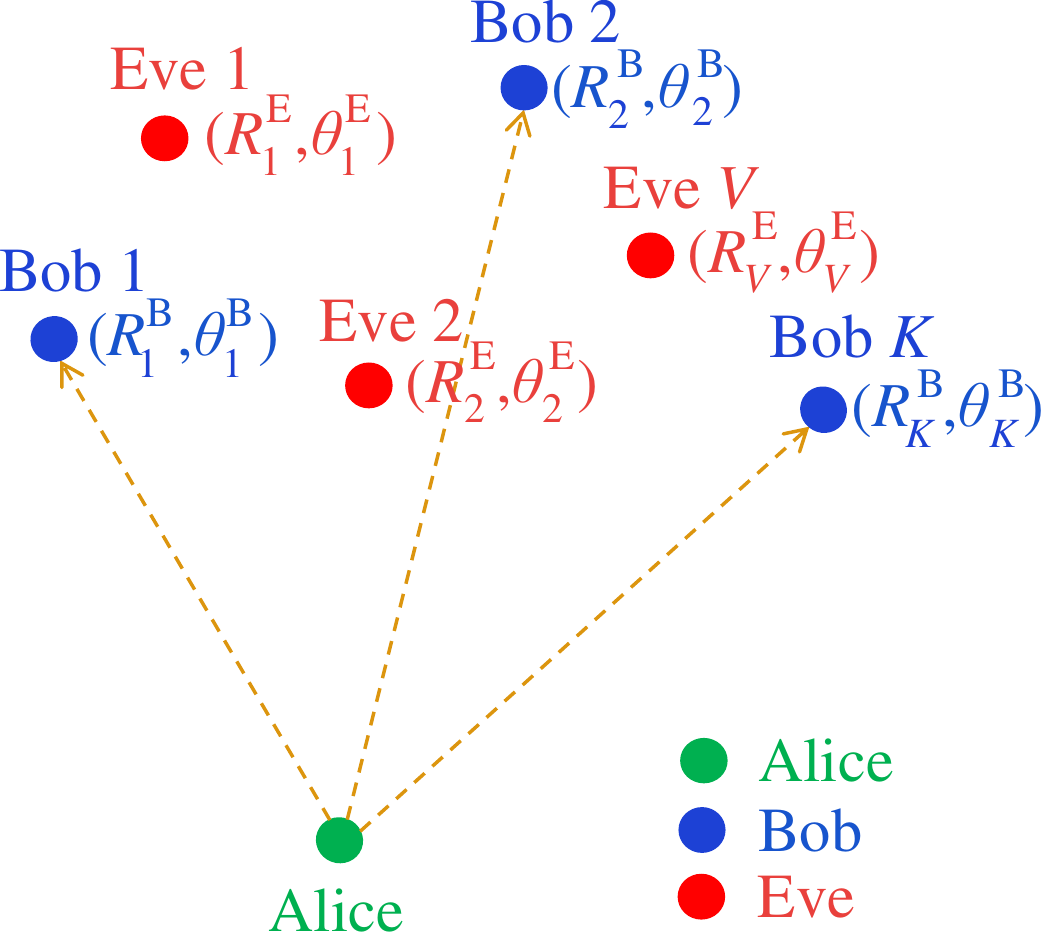}
\caption{The scenario of the proposed power-efficient multi-beam WFRFT-DM scheme with independent receivers.}
\end{figure}

\section{Power-efficient Multi-beam WFRFT-DM Scheme with Independent Receivers}

In this section, we consider another multi-beam DM scenario where the Alice is trying to transmit confidential information to the $K$ independent Bobs located at different places in free space, as shown in Fig. 4. As assumed in Section III, there are also $V$ passive Eves.

\subsection{The Architecture of Alice with Independent Receivers}

As shown in Fig. 5, the Alice in the independent case employs the same symmetrical multi-carrier FDA architecture as Section III. The baseband processing of Alice, however, has to be redesigned because the $K$ Bobs are unable to share information with each other.

{

In the independent case, the WFRFT is operated separately on the data path for each Bob. For the $k$-th data path ($k=1,2,\cdots,K$), a transmission period includes $Q_k$ baseband symbol periods, and $Q_k$ is also the sequence length of WFRFT for the $k$-th Bob.

For the $k$-th data path, the $Q_k$-length pre-transmitting symbol vector can be expressed as ${\bf{s}}_{k}=\left[s_{k,1}~s_{k,2}~\cdots~s_{k,Q_k}\right]^{\rm{T}}$, where $s_{k,q_k}\in\Omega_{k}$ and ${\mathbb{E}}(|s_{k,q_k}|^2)=1$ ($q_k=1,2,\cdots,Q_k$) . Here, $\Omega_{k}$ is the alphabet of baseband symbols for the $k$-th Bob with size $|\Omega_{k}|=M_k$. Then the $Q_k$-length pre-transmitting symbol vector ${\bf{s}}_{k}$ is sent to the WFRFT transformer with the parameter $\alpha_k^{\rm{A}}$, which yields
\begin{subequations}
\label{eq_u_k}
\begin{align}
\tilde{\bf{u}}_{k} &= {\mathscr{F}}^{\alpha_{k}^{\rm{A}}}({\bf{s}}_{k})=\omega_{0,k} {\bf{s}}_{k}+\omega_{1,k} {\mathop{\bf{s}}\limits^{.}}_{k} + \omega_{2,k} {\mathop{\bf{s}}\limits^{..}}_{k} + \omega_{3,k} {\mathop{\bf{s}}\limits^{...}}_{k} \\
&\triangleq\left[{\tilde u}_{k,1}~{\tilde u}_{k,2}~\cdots~{\tilde u}_{k,Q_k}\right]^{\rm{T}}
\end{align}
\end{subequations}
where ${\mathop{\bf{s}}\limits^{.}}_{k}={\mathscr{D}}({\bf{s}}_{k})$, ${\mathop{\bf{s}}\limits^{..}}_{k}={\mathscr{D}}({\mathop{\bf{s}}\limits^{.}}_{k})$, ${\mathop{\bf{s}}\limits^{...}}_{k}={\mathscr{D}}({\mathop{\bf{s}}\limits^{..}}_{k})$, and ${\bf{s}}_{k}={\mathscr{D}}({\mathop{\bf{s}}\limits^{...}}_{k})$. $\omega_{i,k}$ ($i=0,1,2,3$) are the weight coefficients for the $k$-th data path, which can be calculated by replacing $\alpha$ with $\alpha_k^{\rm{A}}$ in (\ref{eq_omega}).

\begin{figure}
\centering
\includegraphics[angle=0,width=0.49\textwidth]{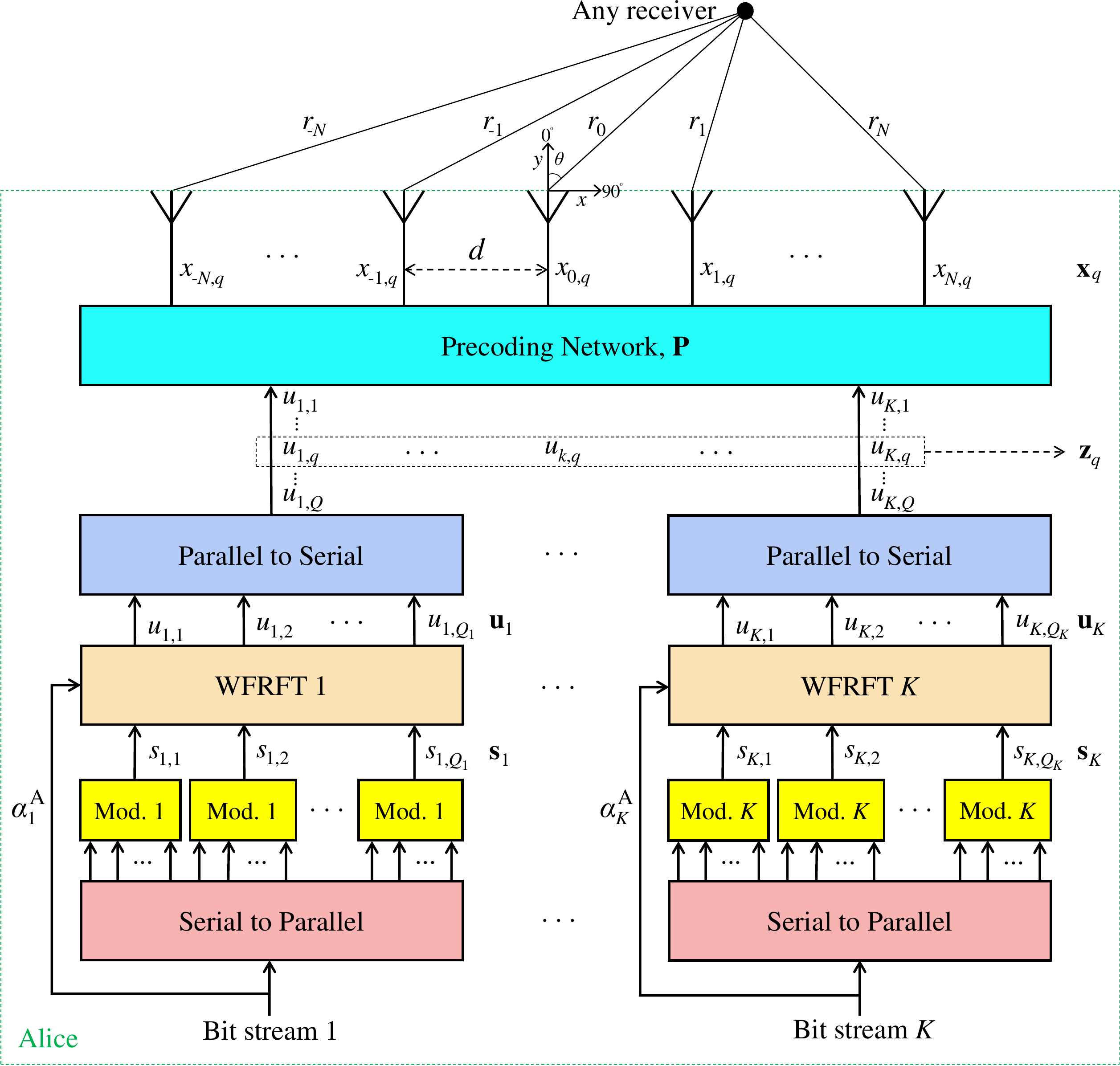}
\caption{{The architecture of Alice for the proposed power-efficient multi-beam WFRFT-DM scheme with independent receivers.}}
\end{figure}

In order to facilitate analysis, we let 
\begin{equation}
\label{eq_Q}
Q=\max \{Q_1,Q_2,\cdots,Q_K\}
\end{equation}
and 
\begin{equation}
\label{eq_U1}
{{\bf{u}}}_{k} = [\tilde{\bf{u}}^{\rm T}_{k}~~\breve{\bf{u}}^{\rm T}_{k}]^{\rm T}
\end{equation}
where $\breve{\bf{u}}_{k} \triangleq \left[{\breve u}_{k,1}~{\breve u}_{k,2}~\cdots~{\breve u}_{k,Q-Q_k}\right]^{\rm{T}}$ is a ($Q-Q_k$)$\times 1$ vector, of which the entries are the symbols after WFRFT in the successive transmission periods for the $k$-th Bob.  Next, a transmitting matrix ${\bf{U}}$ with size $Q\times K$ can be obtained by combining the $K$ symbol vectors $\{{\bf{u}}_{k}\}$, i.e., 
\begin{subequations}
\label{eq_U}
\begin{align}
{\bf{U}}&= \left[{\bf{u}}_{1}~{\bf{u}}_{2}~\cdots~{\bf{u}}_{K}\right]\\
& = \left[ 
\begin{array}{cccc}
{\tilde u}_{1,1} & {\tilde u}_{2,1} & \cdots & {\tilde u}_{K,1}\\
   \vdots &    \vdots & \ddots &    \vdots\\
{\tilde u}_{1,Q_1} & {\tilde u}_{2,Q_2} & \cdots & {\tilde u}_{K,Q_K}\\
{\breve u}_{1,1} & {\breve u}_{2,1} & \cdots & {\breve u}_{K,1}\\
\vdots & \vdots & \ddots & \vdots\\
{\breve u}_{1,Q-Q_1} & {\breve u}_{2,Q-Q_2} & \cdots & {\breve u}_{K,Q-Q_K}
\end{array} 
\right]\\
& \triangleq \left[ 
\begin{array}{cccc}
{u}_{1,1} & {u}_{2,1} & \cdots & {u}_{K,1}\\
{u}_{1,2} & {u}_{2,2} & \cdots & {u}_{K,2}\\
   \vdots &    \vdots & \ddots &    \vdots\\
{u}_{1,Q} & {u}_{2,Q} & \cdots & {u}_{K,Q}
\end{array} 
\right]
\triangleq 
\left[ 
\begin{array}{c}
{\bf{z}}_{1}^{\rm{T}}\\
{\bf{z}}_{2}^{\rm{T}}\\
   \vdots\\
{\bf{z}}_{Q}^{\rm{T}}
\end{array} 
\right]
\end{align}
\end{subequations}
where ${\bf{z}}_{q}=\left[{{u}}_{1,q}~{{u}}_{2,q}~\cdots~{{u}}_{K,q}\right]^{\rm{T}}$ ($q=1,2,\cdots,Q$) is the transpose of the $q$-th row vector of ${\bf{U}}$. Then ${\bf{z}}_{q}$ is utilized to generate the transmitting data for the $2N+1$ antenna elements during the $q$-th baseband symbol period after precoding. It is worth noting that owing to the different lengths of WFRFT sequences for different Bobs, the symbols in (\ref{eq_U}) may include the symbols in one or multiple transmission periods for different Bobs.

\begin{figure}
\centering
\includegraphics[angle=0,width=0.35\textwidth]{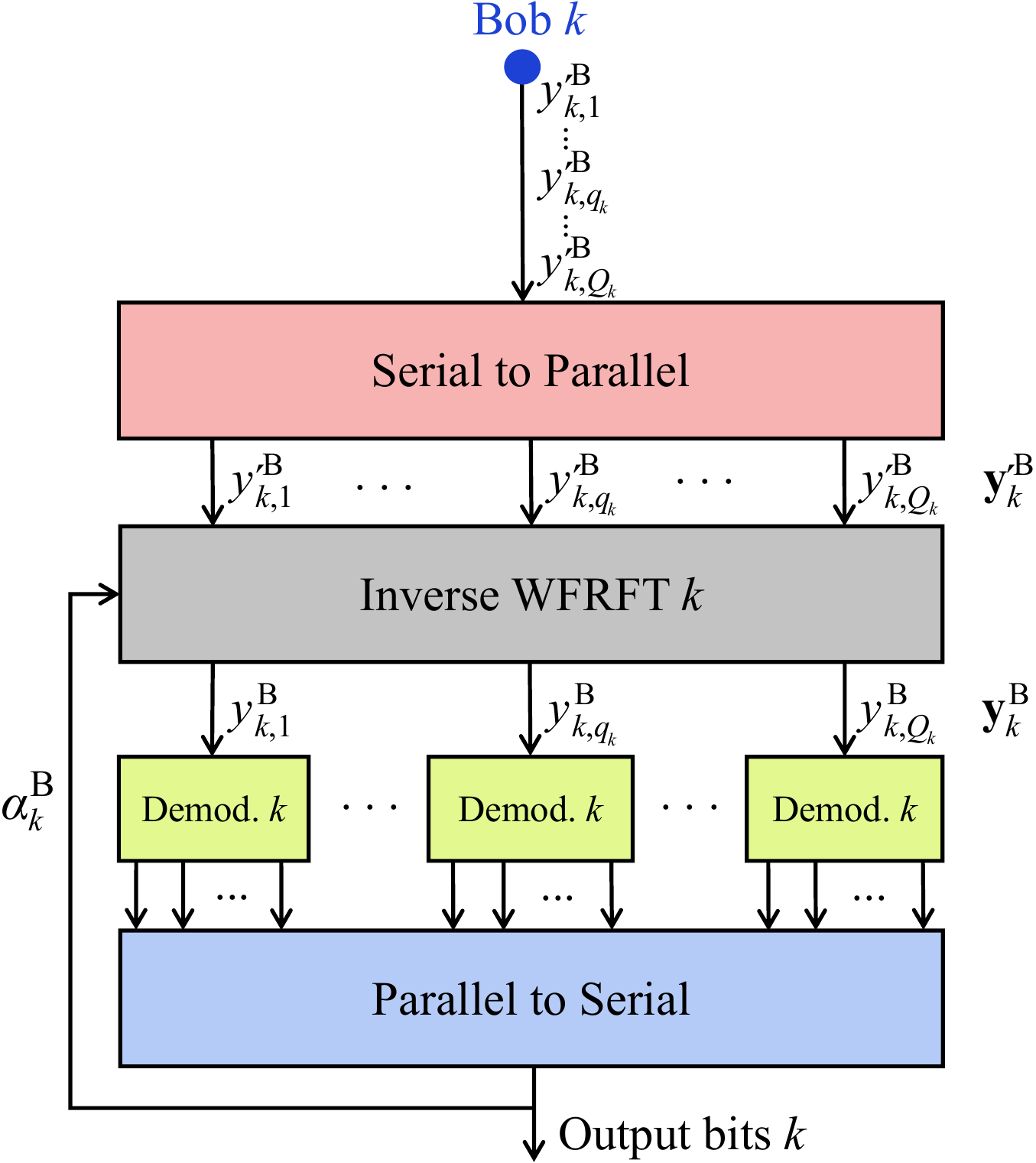}
\caption{{The architecture of the $k$-th Bob for the proposed power-efficient multi-beam WFRFT-DM scheme with independent receivers.}}
\end{figure}

Since the spatial locations of the $K$ independent Bobs are assumed to be prior known for Alice, the precoding matrix is designed the same as (\ref{eq_P_normalized}) in the cooperative case. After precoding, the radiating data for the $2N+1$ antenna elements during the $q$-th symbol period can be produced by 
\begin{equation}
\label{eq_x_q}
{\bf{x}}_{q}=\sqrt{P_s}{\bf{P}}{\bf{z}}_{q}
\end{equation}

\subsection{The Architecture of Bob with Independent Receivers}

To match the architecture of Alice, the architecture of the $k$-th Bob in the independent case is redesigned in Fig. 6.

After passing the LoS channel in free space, the received signal of the $k$-th Bob during the $q_k$-th symbol period can be written as
\begin{subequations}
\label{eq_y_k_q_B1_1}
\begin{align}
{{y}'}_{k,q_k}^{\rm{B}} &={{y}'}_{q_k}(R_{k}^{\rm{B}},\theta_{k}^{\rm{B}})={\bf{h}}^{\rm{H}}(R_{k}^{\rm{B}},\theta_{k}^{\rm{B}}){\bf{x}}_{q_k} + {{\xi}'}_{k,q_k}^{\rm{B}} \\
&=\sqrt{P_s}{\bf{h}}^{\rm{H}}(R_{k}^{\rm{B}},\theta_{k}^{\rm{B}}){\bf{P}}{\bf{z}}_{q_k} + {{\xi}'}_{k,q_k}^{\rm{B}}
\end{align}
\end{subequations}
where ${\xi'}_{k,q_k}^{\rm{B}} \sim {\cal CN}(0,{\sigma_{\xi^{\rm{B}}} ^2})$ is the complex AWGN with zero mean and variance ${\sigma_{\xi^{\rm{B}}} ^2}$. 

According to the normalization property in (\ref{eq_P_H_relation}), the received signal in (\ref{eq_y_k_q_B1_1}) can be further simplified as 
\begin{equation}
\label{eq_y_k_q_B1_2}
{{y}'}_{k,q_k}^{\rm{B}}=\sqrt{P_s}{\bf{I}}_{k}{\bf{z}}_{q_k} + {{\xi}'}_{k,q_k}^{\rm{B}}=\sqrt{P_s}{\tilde u}_{k,q_k} + {{\xi}'}_{k,q_k}^{\rm{B}}
\end{equation}
where ${\bf{I}}_{k}$ is the $k$-th row of the identity matrix ${\bf{I}}_{K}$.

After receiving all of the transmitted signals during the $Q_k$ symbol periods for the $k$-th Bob, the received signal vector can be formed as
\begin{subequations}
\label{eq_Y_k_B1}
\begin{align}
{{\bf{y}}'}_{k}^{\rm{B}}&={{\bf{y}}'}(R_{k}^{\rm{B}},\theta_{k}^{\rm{B}})=\left[{{y}'}_{k,1}^{\rm{B}}~{{y}'}_{k,2}^{\rm{B}}~\cdots~{{y}'}_{k,Q_k}^{\rm{B}}\right]^{\rm{T}}\\
&=\sqrt{P_s}\left[{\tilde u}_{k,1}~{\tilde u}_{k,2}~\cdots~{\tilde u}_{k,Q_k}\right]^{\rm{T}} + {{\boldsymbol{\xi}}'}_{k}^{\rm{B}}=\sqrt{P_s}{\tilde {\bf u}}_{k} + {{\boldsymbol{\xi}}'}_{k}^{\rm{B}}
\end{align}
\end{subequations}
where ${{\boldsymbol{\xi}}'}_{k}^{\rm{B}} = \left[{{\xi}'}_{k,1}^{\rm{B}}~{{\xi}'}_{k,2}^{\rm{B}}~\cdots~{{\xi}'}_{k,Q_k}^{\rm{B}}\right]^{\rm{T}}$ is a complex AWGN vector following $ {\cal CN}({\bf{0}}_{Q_k\times 1},{\sigma_{\xi^{\rm{B}}} ^2}{\bf{I}}_{Q_k})$.

Afterwards, the received $Q_k$-length signal vector ${{\bf{y}}'}_{k}^{\rm{B}}$ is sent to the inverse WFRFT operator with the parameter
\begin{equation}
\label{eq_alpha_k_B}
{\alpha}_{k}^{\rm{B}}=-{\alpha}_{k}^{\rm{A}}
\end{equation}
which yields
\begin{subequations}
\label{eq_Y_k_B2}
\begin{align}
{\bf{y}}_{k}^{\rm{B}}&={\bf{y}}(R_{k}^{\rm{B}},\theta_{k}^{\rm{B}})={\mathscr{F}}^{{\alpha}_{k}^{\rm{B}}}({{\bf{y}}'}_{k}^{\rm{B}})={\mathscr{F}}^{{\alpha}_{k}^{\rm{B}}}(\sqrt{P_s}{\tilde {\bf u}}_{k} + {{\boldsymbol{\xi}}'}_{k}^{\rm{B}})\\
&=\sqrt{P_s}{\mathscr{F}}^{-{\alpha}_{k}^{\rm{A}}}{\mathscr{F}}^{{\alpha}_{k}^{\rm{A}}}({\bf{s}}_{k}) + {\mathscr{F}}^{{\alpha}_{k}^{\rm{B}}}({{\boldsymbol{\xi}}'}_{k}^{\rm{B}}) =\sqrt{P_s}{\bf{s}}_{k} + {\boldsymbol{\xi}}_{k}^{\rm{B}}
\end{align}
\end{subequations}
where ${\boldsymbol{\xi}}_{k}^{\rm{B}} = \left[{\xi}_{k,1}^{\rm{B}}~{\xi}_{k,2}^{\rm{B}}~\cdots~{\xi}_{k,Q_k}^{\rm{B}}\right]^{\rm{T}} \sim {\cal CN}({\bf{0}}_{Q_k\times 1},{\sigma_{\xi^{\rm{B}}} ^2}{\bf{I}}_{Q_k})$ is the AWGN after WFRFT, which has the same distribution characteristics as ${{\boldsymbol{\xi}}'}_{k}^{\rm{B}}$ \cite{Mei_Research_on_4WFRFT}.

Therefore, the received signal of the $k$-th Bob during the $q_k$-th symbol period is
\begin{equation}
\label{eq_y_k_q_B2}
{y}_{k,q_k}^{\rm{B}} = {y}_{q_k}(R_{k}^{\rm{B}},\theta_{k}^{\rm{B}})=\sqrt{P_s}{s}_{k,q_k} + {\xi}_{k,q_k}^{\rm{B}}
\end{equation}
From (\ref{eq_y_k_q_B2}), the $k$-th Bob can easily recover the $q_k$-th confidential information transmitted from Alice.

}

{
\subsection{Eves' Received Signals with Independent Receivers}
}

For the $v$-th Eve ($v=1,2,\cdots,V$) located at $({R_v^{\rm{E}}},\theta_v^{\rm{E}})$, its received signal during the $q$-th symbol period  can be expressed as
\begin{subequations}
\label{eq_y_v_q_E1}
\begin{align}
{y}_{v,q}^{\rm{E}}&={y}_{q}({R_v^{\rm{E}}},\theta_v^{\rm{E}})={\bf{h}}^{\rm{H}}({R_v^{\rm{E}}},\theta_v^{\rm{E}}){\bf{x}}_{q} + {{\xi}}_{v,q}^{\rm{E}}\\
&=\sqrt{P_s}{\bf{h}}^{\rm{H}}({R_v^{\rm{E}}},\theta_v^{\rm{E}}){\bf{P}}{\bf{z}}_{q} + {{\xi}}_{v,q}^{\rm{E}}
\end{align}
\end{subequations}
where ${\xi}_{v,q}^{\rm{E}} \sim {\cal CN}(0,{\sigma_{\xi^{\rm{E}}} ^2})$ is the complex AWGN with zero mean and variance ${\sigma_{\xi^{\rm{E}}} ^2}$. Noting that the Eves cannot recover the confidential information from (\ref{eq_y_v_q_E1}), because the steering vectors of Eves are not normalized with $\bf{P}$ and they are also ignorant of the length of symbol sequence for WFRFT.

Here, we consider the scenario where the Eves are ignorant of the WFRFT parameters shared between Alice and Bobs but they can obtain the length of symbol sequence for WFRFT. After receiving all the signals during the $Q$ symbol periods, the $v$-th Eve can combine the received signals in (\ref{eq_y_v_q_E1}) as a vector, i.e.,  
\begin{subequations}
\label{eq_Y_E1}
\begin{align}
{{\bf{y}}}_v^{\rm{E}}&={{\bf{y}}}({R_v^{\rm{E}}},\theta_v^{\rm{E}})=\left[{{y}}_{v,1}^{\rm{E}}~{{y}}_{v,2}^{\rm{E}}~\cdots~{{y}}_{v,Q}^{\rm{E}}\right]^{\rm{T}} \\
&= {\left\{ {\sqrt {{P_s}} {{\bf{h}}^{\rm{H}}}({R_v^{\rm{E}}},\theta _v^{\rm{E}}){\bf{P}}\left[ {{{\bf{z}}_1}~{{\bf{z}}_2}~ \cdots ~{{\bf{z}}_Q}} \right]} \right\}^{\rm{T}}}  + {\boldsymbol{\xi}}_{v}^{\rm{E}}\\
&=\sqrt {{P_s}} \left[ 
{\begin{array}{*{20}{c}}
{{\bf{z}}_1^{\rm{T}}}\\
{{\bf{z}}_2^{\rm{T}}}\\
 \vdots \\
{{\bf{z}}_Q^{\rm{T}}}
\end{array}}
 \right]{{\bf{P}}^{\rm{T}}}{{\bf{h}}^ *}({R_v^{\rm{E}}},\theta _v^{\rm{E}}) + {\boldsymbol{\xi}}_{v}^{\rm{E}} \\
 &= \sqrt {{P_s}} [{{\bf{u}}_1}~{{\bf{u}}_2}~ \cdots ~{{\bf{u}}_K}]{{\bf{P}}^{\rm{T}}}{{\bf{h}}^ * }({R_v^{\rm{E}}},\theta _v^{\rm{E}}) + {\boldsymbol{\xi}}_{v}^{\rm{E}}
\end{align}
\end{subequations}
where ${{\boldsymbol{\xi}}_{v}^{\rm{E}}}= \left[{\xi}_{v,1}^{\rm{E}}~{\xi}_{v,2}^{\rm{E}}~\cdots~{\xi}_{v,Q}^{\rm{E}} \right]^{\rm{T}}$ is the AWGN vector.

Letting 
\begin{equation}
\label{eq_Rho_v}
{\boldsymbol{\varrho}}_{v} = {{\bf{P}}^{\rm{T}}}{{\bf{h}}^ {*} }({R_v^{\rm{E}}},\theta _v^{\rm{E}}) =[{\varrho}_{v,1}~{\varrho}_{v,2}~\cdots~{\varrho}_{v,K}]^{\rm{T}} 
\end{equation}
then the received signal vector in (\ref{eq_Y_E1}) can be simplified into (\ref{eq_Y_E2}) in the next page, where ${\boldsymbol{\eta}}_{k'} = \omega_{1,k'} {\mathop{\bf{s}}\limits^{.}}_{k'} + \omega_{2,k'} {\mathop{\bf{s}}\limits^{..}}_{k'} + \omega_{3,k'} {\mathop{\bf{s}}\limits^{...}}_{k'} = [{\eta}_{k',1}~{\eta}_{k',2}~\cdots~{\eta}_{k',Q}]^{\rm{T}}$, $k'=1,2,\cdots,K$, is the equivalent AN with variance $\sigma_{\eta_{k'}}^2=1-|\omega_{0,k'}|^2$ \cite{Fang_PLS_Cooperation_WFRFT2}.

\newcounter{cnt4}
\setcounter{cnt4}{\value{equation}}
\setcounter{equation}{48}
\begin{figure*}[t]
\begin{subequations}
\label{eq_Y_E2}
\begin{align}
{{\bf{y}}}_{v}^{\rm{E}} &= \sqrt {{P_s}} [{{\bf{u}}_1}~{{\bf{u}}_2}~ \cdots ~{{\bf{u}}_K}]{\boldsymbol{\varrho}}_{v} + {\boldsymbol{\xi}}_{v}^{\rm{E}} = \sqrt {{P_s}} \sum\limits_{k'=1}^{K}{{\varrho}_{v,k'}}{{\bf{u}}_{k'}} + {\boldsymbol{\xi}}_{v}^{\rm{E}} = \sqrt {{P_s}} \sum\limits_{k'=1}^{K}{{\varrho}_{v,k'}}{\mathscr{F}}^{\alpha_{k'}^{\rm{A}}}({\bf{s}}_{k'}) + {\boldsymbol{\xi}}_{v}^{\rm{E}}\\
&=\sqrt {{P_s}} \sum\limits_{k'=1}^{K}{{\varrho}_{v,k'}}(\omega_{0,k'} {\bf{s}}_{k'}+\omega_{1,k'} {\mathop{\bf{s}}\limits^{.}}_{k'} + \omega_{2,k'} {\mathop{\bf{s}}\limits^{..}}_{k'} + \omega_{3,k'} {\mathop{\bf{s}}\limits^{...}}_{k'}) + {\boldsymbol{\xi}}_{v}^{\rm{E}}\\
&=\underbrace{\sqrt {{P_s}} \sum\limits_{k'=1}^{K}{{\varrho}_{v,k'}} \omega_{0,k'} {\bf{s}}_{k'}}_{{\rm{Mixed~Signal}}} + \underbrace{\sqrt {{P_s}} \sum\limits_{k'=1}^{K}{{\varrho}_{v,k'}} {\boldsymbol{\eta}}_{k'}}_{{\rm{Equivalent~AN}}} + \underbrace{ {\boldsymbol{\xi}}_{v}^{\rm{E}}}_{{\rm{AWGN}}}
\end{align}
\end{subequations}
\hrulefill
\end{figure*}
\setcounter{equation}{\value{cnt4}}
\setcounter{equation}{49}

It is worth noting that the first part of (\ref{eq_Y_E2}c) is the mixed signal containing all the $K$ Bobs' signals. Specifically, if the $v$-th Eve is meant to eavesdrop the $k$-th ($k=1,2,\cdots,K$) Bob's information, the received signal in (\ref{eq_Y_E2}) can be further reorganized as
\begin{subequations}
\label{eq_Y_E3}
\begin{align}
{\bf{y}}_{v,k}^{\rm{E}} & = \underbrace{\sqrt {{P_s}} {{\varrho}_{v,k}} \omega_{0,k} {\bf{s}}_{k}}_{{\rm{Distorted~Signal}}} 
+ \underbrace{\sqrt {{P_s}} \sum\limits_{k'\neq k}{{\varrho}_{v,k'}} \omega_{0,k'} {\bf{s}}_{k'}}_{{\rm{Equivalent~Mixed~Noise}}} \\
&~~~+ \underbrace{\sqrt {{P_s}} \sum\limits_{k'=1}^{K}{{\varrho}_{v,k'}} {\boldsymbol{\eta}}_{k'}}_{{\rm{Equivalent~AN}}} 
+ \underbrace{ {\boldsymbol{\xi}}_{v}^{\rm{E}}}_{{\rm{AWGN}}}
\end{align}
\end{subequations}
 
  Furthermore, when the $v$-th Eve is trying to eavesdrop the $k$-th Bob's information, the received signal during the $q$-th symbol period can be expressed as 
\begin{subequations}
\label{eq_y_q_E2}
\begin{align}
{y}_{v,k,q}^{\rm{E}} &= \underbrace{\sqrt {{P_s}} {{\varrho}_{v,k}} \omega_{0,k} {s}_{k,q}}_{{\rm{Distorted~Signal}}} 
+ \underbrace{\sqrt {{P_s}} \sum\limits_{k'\neq k}{{\varrho}_{v,k'}} \omega_{0,k'} {s}_{k',q}}_{{\rm{Equivalent~Mixed~Noise}}} \\
& ~~~+ \underbrace{\sqrt {{P_s}} \sum\limits_{k'=1}^{K}{{\varrho}_{v,k'}} {\eta}_{k',q}}_{{\rm{Equivalent~AN}}} 
+ \underbrace{ {\xi}_{v,q}^{\rm{E}}}_{{\rm{AWGN}}}
\end{align}
\end{subequations}

From (\ref{eq_Y_E3}) and (\ref{eq_y_q_E2}), it is noting that the received signal of the $v$-th Eve when eavesdropping the $k$-th Bob's information is composed by four components. The first is the $k$-th Bob's distorted signal, the second is the equivalent mixed noise from other Bobs' signals, the third is the equivalent AN caused by WFRFT, and the last is the AWGN. 

Analogous to the proposed multi-beam WFRFT-DM scheme with cooperative receivers, there is also no actual AN inserted in the transmitting signals for the proposed multi-beam WFRFT-DM scheme with independent receivers. Despite no actual AN inserted, the equivalent AN can be effectively produced for Eves by means of the application of WFRFT. Additionally, due to the existence of the equivalent mixed noise the equivalent AN, the transmission security (\emph{neighbor security}) can also be guaranteed with close or identical locations of Eves and Bobs.

\section{Performance Analysis}

BER, secrecy rate, and robustness are three important metrics to measure the performance of DM systems \cite{Ding_BER_DM}\cite{Ding_Metrics_DM}. In this section, we will analyze the BER, secrecy rate, and robustness of the proposed power-efficient multi-beam WFRFT-DM schemes. {The comparisons among different multi-beam DM transmission schemes are also provided. } Moreover, some practical issues to implement the proposed multi-beam WFRFT-DM schemes are discussed as well.

\subsection{BER}
Without loss of generality, we assume $\sigma_{{\xi}^{\rm{B}}}^{2}=\sigma_{{\xi}^{\rm{E}}}^{2}=\sigma_{{\xi}}^{2}$ in this paper. Since the baseband symbols are normalized, the signal-to-noise ratio (SNR) $\gamma$ can be expressed as 
\begin{equation}
\label{eq_SNR}
\gamma = \frac{ P_{s} } {\sigma_{{\xi}^{\rm{B}}}^{2}}= \frac{ P_{s} } {\sigma_{{\xi}^{\rm{E}}}^{2}}=\frac{ P_{s} } {\sigma_{{\xi}}^{2}}
\end{equation}

By contrast, the SNR of the conventional multi-beam AN-DM scheme can be written as
\begin{equation}
\label{eq_SNR_AN}
\gamma^{\rm{AN}} = \frac{ \beta_{1}^{2}P_{s} } {\sigma_{{\xi}}^{2}} = \beta_{1}^{2} \gamma
\end{equation}
where $\beta_{1}$ ($0<\beta_{1}<1$) is the power splitting factor for the useful signal.

It can be observed from (\ref{eq_y_r_theta_k_B2}) and (\ref{eq_y_k_q_B2}) that the received signals of Bobs are simply the sum of useful signals and AWGN. Therefore, if we take the modulation mode of $M$-PSK into consideration, the average SNR per bit $\gamma _b$ can be calculated by 
\begin{equation}
\label{eq_EbNo}
\gamma _b=\frac{P_s E_s}{ \sigma_{\xi}^{2} \log _{2} M} = \frac{\gamma}{\log _{2} M} 
\end{equation}
where $E_s$ is the average normalized energy per symbol, which is set as the normalized value in this paper, i.e., $E_s = \mathbb{E}(|{s_{k}}|^2)=1$.

Using (\ref{eq_EbNo}), the BER of the proposed multi-beam WFRFT-DM schemes can be calculated by \cite{Proakis_Digital_Commu}
\begin{subequations}
\label{eq_BER1}
\begin{align}
BER &\approx \frac{2}{\log_2 M}Q\left(\sqrt{2\gamma_b {\log_2 M}}\sin\left(\frac{\pi}{M}\right)\right)\\
& = \frac{2}{\log_2 M}Q\left(\sqrt{2\gamma }\sin\left(\frac{\pi}{M}\right)\right)
\end{align}
\end{subequations}
where $Q(t) = {1 \mathord{\left/
 {\vphantom {1 {\sqrt {2\pi } }}} \right.
 \kern-\nulldelimiterspace} {\sqrt {2\pi } }}\int_t^\infty  {\exp \{  - {{{t^2}} \mathord{\left/
 {\vphantom {{{t^2}} 2}} \right.
 \kern-\nulldelimiterspace} 2}\} } dt$ is the tail distribution function of the standard normal distribution.

\subsection{Secrecy Rate}

$\bullet$ Cooperative WFRFT-DM Scheme

According to (\ref{eq_y_r_theta_k_B2}), the signal to interference-plus-noise ratio (SINR) of the $k$-th Bob in the cooperative case can be expressed as
\begin{equation}
\label{eq_SINR_k_B_Coop}
\lambda _{k}^{{\rm{B}},\rm{Co}} = \frac{ P_{s} } {\sigma_{{\xi}^{\rm{B}}}^{2}}= \gamma
\end{equation}

Using (\ref{eq_SINR_k_B_Coop}), the achievable rate of the link from Alice to the $k$-th Bob in the cooperative case can be calculated by \cite{Ji_Fading_FDA_DM}
\begin{equation}
\label{eq_R_k_B_Coop}
{\Gamma_{k}^{{\rm{B}},\rm{Co}}}=\log_{2}\left(1+\lambda _{k}^{{\rm{B}},\rm{Co}}\right)
\end{equation} 

According to (\ref{eq_y_r_theta_v_E}), the SINR of the $v$-th Eve in the cooperative case can be calculated by
\begin{equation}
\label{eq_SINR_v_E_Coop}
\lambda _{v}^{{\rm{E}},\rm{Co}} = \frac{ P_{s} |\omega_{0}|^{2} \mathbf{h}^{\rm{H}}(R_{v}^{\rm{E}},\theta_{v}^{\rm{E}}) \mathbf{P} \mathbf{P}^{\rm{H}} \mathbf{h}(R_{v}^{\rm{E}},\theta_{v}^{\rm{E}})} {P_{s} \mathbf{h}^{\rm{H}}(R_{v}^{\rm{E}},\theta_{v}^{\rm{E}}) \mathbf{P} \mathbf{P}^{\rm{H}} \mathbf{h}(R_{v}^{\rm{E}},\theta_{v}^{\rm{E}})\sigma_{\eta}^{2} + \sigma_{{\xi}^{\rm{E}}}^{2} }
\end{equation}

By means of (\ref{eq_SINR_v_E_Coop}), we can obtain the achievable rate of the link from Alice to the $v$-th Eve in the cooperative case as
\begin{equation}
\label{eq_R_v_E_Coop}
{\Gamma_{v}^{{\rm{E}},\rm{Co}}}=\log_{2}\left(1+\lambda _{v}^{{\rm{E}},\rm{Co}}\right)
\end{equation}

Therefore, the secrecy rate of the proposed multi-beam WFRFT-DM scheme with cooperative receivers can be defined as \cite{Ji_Fading_FDA_DM}
\begin{equation}
\label{eq_secrecy_rate_Coop}
{\Gamma_s^{\rm{Co}}}=\max\limits_{k\in\left\{1,\cdots,K \right\}}\left[\min\limits_{v\in\left\{1,\cdots,V \right\}}\left({\Gamma_{k}^{{\rm{B}},\rm{Co}}-\Gamma_{v}^{{\rm{E}},\rm{Co}}}\right)  \right]^{+} 
\end{equation}
where $[\cdot]^{+}=\max\{0,\cdot\}$.

$\bullet$ Independent WFRFT-DM Scheme

Similarly, according to (\ref{eq_y_k_q_B2}), the SINR of the $k$-th Bob in the independent case can be calculated by
\begin{equation}
\label{eq_SINR_k_q_B_Inde}
\lambda _{k}^{{\rm{B}},\rm{In}} = \frac{ P_{s} } {\sigma_{{\xi}^{\rm{B}}}^{2}} =\gamma
\end{equation}

The achievable rate of the link from Alice to the $k$-th Bob in the independent case is given as
\begin{equation}
\label{eq_R_k_B_Inde}
{\Gamma_{k}^{{\rm{B}},\rm{In}}}=\log_{2}\left(1+\lambda _{k}^{{\rm{B}},\rm{In}}\right)
\end{equation}

According to (\ref{eq_Y_E3}) and (\ref{eq_y_q_E2}), the SINR of the $v$-th Eve when eavesdropping the $k$-th Bob's information in the independent case can be calculated by
\begin{equation}
\label{eq_SINR_k_q_E_Inde}
\lambda _{v,k}^{{\rm{E}},\rm{In}} =\frac{ P_{s} |\varrho_{v,k}|^{2} |\omega_{0,k}|^{2}}
{P_{s}\sum\limits_{k'\neq k}|{{\varrho}_{v,k'}}|^{2} |\omega_{0,k'}|^{2} 
+ P_{s}\sum\limits_{k'=1}^{K} |{{\varrho}_{v,k'}}|^{2} \sigma_{{\eta}_{k'}}^{2}
 + \sigma_{{\xi}^{\rm{E}}}^{2}}
\end{equation}

Moreover, the achievable rate of the link from Alice to the $v$-th Eve to when eavesdropping the $k$-th Bob's information in the independent case can be expressed as
\begin{equation}
\label{eq_R_v_E}
{\Gamma_{v,k}^{{\rm{E}},\rm{In}}}=\log_{2}\left(1+\lambda _{v,k}^{{\rm{E}},\rm{In}}\right)
\end{equation}

Therefore, the secrecy rate of the proposed multi-beam WFRFT-DM scheme with independent receivers can be defined as \cite{Ji_Fading_FDA_DM}
\begin{equation}
\label{eq_secrecy_rate_Inde}
\begin{aligned}
{\Gamma_s^{\rm{In}}}=\max\limits_{k\in\left\{1,\cdots,K \right\}}\left[\min\limits_{v\in\left\{1,\cdots,V \right\}}\left({\Gamma_{k}^{\rm{B},\rm{In}}}-\max\limits_{k'\in\left\{1,\cdots,K \right\}} {\Gamma_{v,k'}^{\rm{E},\rm{In}}}\right)  \right]^{+}
\end{aligned}
\end{equation}

\subsection{Robustness}

In this section, we will investigate the robustness of the proposed WFRFT-DM schemes including the impacts of imperfect estimation of Bobs' locations and mismatched WFRFT parameters.

$\bullet$ Imperfect Estimation of Bobs' Locations

In the proposed multi-beam WFRFT-DM schemes, the locations of Bobs are assumed to be prior known for Alice. Practically, the locations of Bobs can be obtained by jointly using direction of arrival (DOA) estimation algorithm such as \cite{Shu_DOA_Estimation} and range estimation algorithm such as \cite{Coluccia_Range_Estimation}. 

We assume the estimated location of the $k$-th Bob is
\begin{equation}
\label{eq_Est_error}
({R_{k}^{\rm{B,Est}}},\theta _{k}^{\rm{B,Est}}) =  ({R_{k}^{\rm{B}}},\theta _{k}^{\rm{B}}) + ({\Delta R_{k}},\Delta \theta_{k})
\end{equation}
where ${\Delta R_{k}}$ and $\Delta \theta_{k}$ are the estimated range and angle errors, respectively.

Noting that the estimation errors only impose impact on the precoding matrix $\bf{P}$, while the architectures of Alice and Bobs remain unchanged. In this case, the precoding matrix with estimation errors should be updated to
\begin{equation}
\label{eq_P_Est}
\begin{aligned}
{\mathbf{P}}^{\rm{Est}}&=\mathbf{H}(\Upsilon^{\rm{B,Est}},\Theta^{\rm{B,Est}})\\
&~~~\cdot\left[\mathbf{H}^{\rm{H}}(\Upsilon^{\rm{B,Est}},\Theta^{\rm{B,Est}})\mathbf{H}(\Upsilon^{\rm{B,Est}},\Theta^{\rm{B,Est}})\right]^{-1}
\end{aligned}
\end{equation} 
where $\mathbf{H}(\Upsilon^{\rm{B,Est}},\Theta^{\rm{B,Est}})$ is the estimated steering matrix composed by the $K$ estimated steering vectors, i.e.,
\begin{equation} 
\label{eq_H_R_Theta_B_Est}
\begin{aligned}
&\mathbf{H}(\Upsilon^{\rm{B,Est}},\Theta^{\rm{B,Est}})=\\
&\left[\mathbf{h}({R_1^{\rm{B,Est}}},\theta_1^{\rm{B,Est}})~\mathbf{h}({R_2^{\rm{B,Est}}},\theta_2^{\rm{B,Est}})~\cdots~\mathbf{h}({R_K^{\rm{B,Est}}},\theta_K^{\rm{B,Est}})\right]
\end{aligned}
\end{equation}

From (\ref{eq_P_Est}), the normalization characteristic in (\ref{eq_P_H_relation}) will be affected when there are estimation errors.  Simulations will be provided in Section VI to demonstrate that the security of the proposed multi-beam WFRFT-DM schemes can also hold as long as the estimation errors of Bobs' locations are in an acceptable range. {In practice, some robust DM synthesis methods \cite{Hu_Robust_DM}\cite{Shu_Robust_MB1}\cite{Shu_Robust_MB2} could be employed to fight against the imperfect estimations of Bobs' locations.}

\begin{table*}[!t]
{
\renewcommand{\arraystretch}{1.2}
\caption{Comparisons for Different Multi-Beam DM Schemes}
\label{table1}
\centering
\begin{tabular}{cccccccc}
\hline
\multirow{3}{*}{\bf{Item}} &\multicolumn{5}{|c}{\bf LoS channel}   &  \multicolumn{2}{|c}{\bf Multi-path channel}                         \\ [0.1ex]
\cline{2-8}
 &\multicolumn{2}{|c}{\bf WFRFT-DM}  & \multicolumn{3}{|c|}{\bf AN-DM}  & \multicolumn{1}{|c|}{\multirow{2}{*}{{\bf AN-DM}$^{\rm [25]}$}} & \\ [0.1ex]
\cline{2-6}
 & \multicolumn{1}{|c|}{{\bf{Coop.}}$^{\star}$} & \multicolumn{1}{|c|}{{\bf{Inde.}}$^{\dagger}$} & \multicolumn{1}{|c|}{{\bf{PA}}$^{\rm [16]-[19]}$ } &\multicolumn{1}{|c|}{{\bf{Ro-PA}}$^{{\ddagger}{\rm [20][21]}} $} & \multicolumn{1}{|c|}{{\bf{FDA}}$^{\rm [22]-[24]}$ } &  & \multicolumn{1}{|c}{\multirow{-2}{*}{{\bf MP-DM}$^{\rm [26][27]}$} }\\ [0.1ex]
\hline
{\bf AN}\protect\footnotemark[1]  & No & No & Yes & Yes & Yes & Yes & No \\ [0.1ex]
{\bf PE}\protect\footnotemark[1]  & Yes & Yes & No & No & No & No & Yes \\ [0.1ex]
{\bf PM}\protect\footnotemark[1]  & Yes & Yes & Yes & Yes & Yes & Yes & Yes \\ [0.1ex]
{\bf OM}\protect\footnotemark[1]  & No & No & Yes & Yes & Yes & Yes & No \\ [0.1ex]
{\bf NS}\protect\footnotemark[1]  & Yes & Yes & No & No & No & No & No \\ [0.1ex]
{\bf RB}\protect\footnotemark[1]  & No & No & No & Yes & No  & No & No \\ [0.1ex]
{\bf CS}\protect\footnotemark[1]  & Yes & No & No & No & No &  No & No \\ [0.1ex]
{\bf RA}\protect\footnotemark[1]  & Yes & Yes & No &  No & Yes & No & Yes\protect\footnotemark[2] \\ [0.1ex]
{\bf IT}\protect\footnotemark[1] & Yes &  Yes & Yes &  Yes & Yes & Yes & Yes \\ [0.1ex]
\hline
\end{tabular}
\\
\footnotesize{$^{\star}$ The cooperative WFRFT-DM scheme; $^{\dagger}$ The independent WFRFT-DM scheme; $^{\ddagger}$ The robust multi-beam AN-DM schemes.}\\
}
\end{table*}
\footnotetext[1]{ { AN: Artificial noise; PE: Power-efficient; PM: Precoding matrix; OM: Orthogonal matrix; NS: Neighbor security (Eve's location is close to or the same as Bob's); RB: Robustness (Imperfect estimations of Bobs' locations); CS: Cooperative sharing; RA: Range-angle dependent; IT: Independent transmission (Different modulations for Bobs).}}
\footnotetext[2]{{ The location-specific coordinated multi-point (CoMP) transmission in [27] requires joint processing of multiple base stations. In this paper, by contrast, only one base station is considered.}}

$\bullet$ Mismatched WFRFT Parameters

For the proposed multi-beam WFRFT-DM schemes, Alice and Bobs have to share the WFRFT parameters, which may actually disagree owing to the imperfect key establishing link. Here we assume the practical WFRFT parameter of Bob is  
\begin{equation}
\label{eq_practical_alpha_B}
\alpha^{\rm{B,Pra}} = -\alpha^{\rm{A}}+\Delta\alpha
\end{equation}
where $\Delta\alpha$ refers to the mismatch parameter between Alice and Bob, which, according to (\ref{eq_alpha_B}), equals zero in the ideal case. Simulations will be provided in the following section to investigate the impact of mismatched WFRFT parameter on the proposed WFRFT-DM schemes.

{

\subsection{Comparisons for Different Multi-beam  DM  Schemes}

Table I fully compares different multi-beam DM schemes in different aspects. Two types of multi-beam DM schemes can be generalized in terms of different channel models. The first is in LoS channels, including the conventional AN-based schemes and the proposed power-efficient non-AN schemes. The second is in multi-path channels, which uses the multi-path nature of channels to create a multi-user interference environment. From Table I, we can generalize the advantages of our proposed WFRFT-DM schemes as follows:

\begin{enumerate}
\item Since no AN is inserted in the transmitting signal, the proposed multi-beam WFRFT-DM schemes are more power-efficient than the conventional multi-beam AN-DM scheme. Moreover, only the precoding matrix is required for the proposed schemes, while both precoding and orthogonal matrices are required for the conventional AN-based scheme.

\item The range-angle dependence is naturally inherited from FDA for the proposed multi-beam WFRFT-DM schemes. Although the schemes in [26] and [27] can achieve location-specific multi-beam DM transmissions over multi-path channels, they will not work in the normalized LoS channel in free space without the assistance of multi-path nature. Especially, the location-specific coordinated multi-point (CoMP) transmission scheme in [27] requires joint processing of multiple base stations. By contrast, to achieve such range-angle dependence, only one transmitter is required in our paper.

\item More importantly, benefited from the inherent security of the WFRFT, the transmission security of the proposed WFRFT-DM schemes can also hold even if Eves' locations are close to or the same as Bobs', which outperform all of other multi-beam DM schemes.\\ 

\end{enumerate}
}

As for the difference between the proposed two multi-beam WFRFT-DM schemes, on the one hand, the cooperation among Bobs is necessary for the cooperative scheme but with only once WFRFT operation; on the other hand, the independent scheme requires more WFRFT operations but with no cooperation required. Therefore, the compromise between the cooperation and the WFRFT operation should be considered when choosing these two different schemes to perform multi-beam DM transmissions.

\subsection{Discussions}

\begin{figure}
\centering
\includegraphics[angle=0,width=0.35\textwidth]{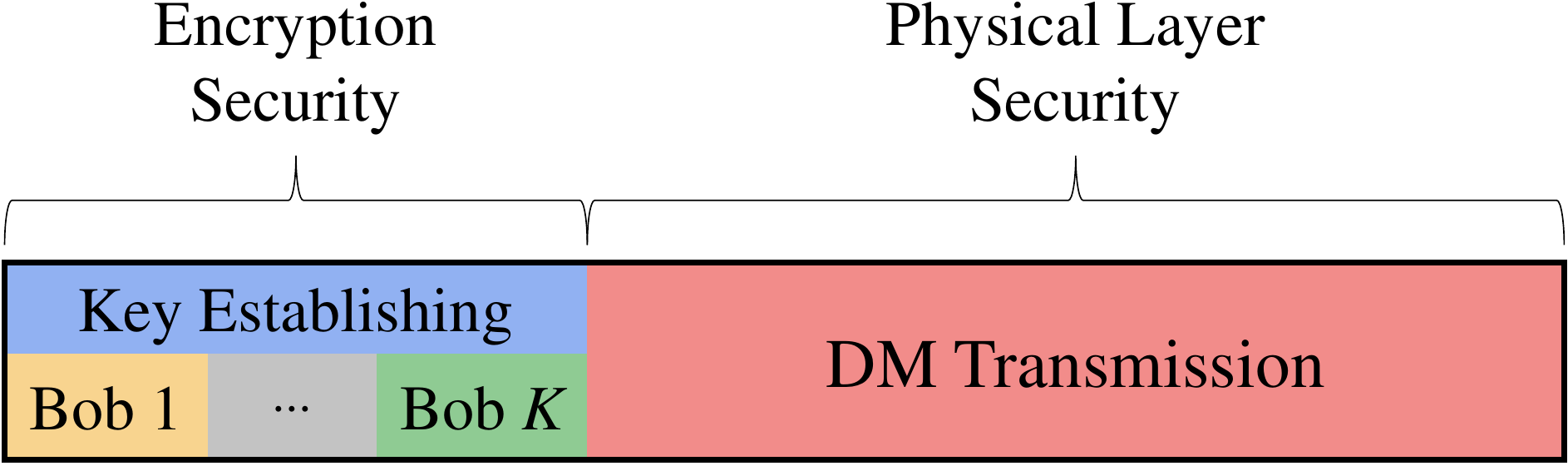}
\caption{Practical WFRFT parameters sharing strategy via key establishing.}
\end{figure}

\begin{table*}[!t]
\renewcommand{\arraystretch}{1.2}
\caption{Simulation Parameters}
\label{table2}
\centering
\begin{tabular}{ll|ll}
\hline
{\bf Parameter} &{\bf Value } & {\bf Parameter} &{\bf Value }\\ [0.1ex]
\hline
Modulation mode & BPSK, QPSK, 8PSK & Location of Bob 1$^{\star}$, $({R_{1}^{\rm{B}}},{\theta _{1}^{\rm{B}}})$ & $(150~{\rm{km}},{50^ \circ })$\\
Central frequency, ${f_0}$ & $10$ GHz  & Location of Bob 2, $({R_{2}^{\rm{B}}},{\theta _{2}^{\rm{B}}})$ & $(180~{\rm{km}},-40^{\circ})$\\
Fixed frequency increment, ${\Delta f}$ & $2$ kHz & Location of Bob 3, $({R_{3}^{\rm{B}}},{\theta _{3}^{\rm{B}}})$ & $(260~{\rm{km}},0^{\circ})$\\
Number of FDA elements, $2N+1$ & $17$ & Location of Eve 1$^{\star}$, $({R_{1}^{\rm{E}}},{\theta _{1}^{\rm{E}}})$ & $(150~{\rm{km}},{50^ \circ })$\\
Number of carriers for each element, $L$ & $7$ & Location of Eve 2, $({R_{2}^{\rm{E}}},{\theta _{2}^{\rm{E}}})$ & $(220~{\rm{km}},{-20^ \circ })$\\
Frequency increment control factor, $p$ & $1$ & WFRFT parameter$^{\dagger}$, $\alpha^{\rm{A}}$ & $0.5$\\
{Total signal power after precoding}, ${P_s}$ & $1$  & WFRFT parameter$^{\ddagger}$, $\alpha_1^{\rm{A}},~\alpha_2^{\rm{A}},~\alpha_3^{\rm{A}}$ & $0.5,~1,~1.5$\\
Number of Bobs, $K$ & $3$ & WFRFT parameter$^{\dagger\ddagger}$, $M_V^{\rm{A}}=M_V^{\rm{B}}$ & $[1~2~3~4]$\\
Number of Eves, $V$ & $2$ & WFRFT parameter$^{\dagger\ddagger}$, $N_V^{\rm{A}}=N_V^{\rm{B}}$ & $[5~6~7~8]$\\
& & {Length of WFRFT}$^{\ddagger}${, $Q_1,Q_2,Q_3$} & {$3,4,5$}\\
\hline
\end{tabular}
\\
\footnotesize{$^{\star}$ Identical location for Bob 1 and Eve 1; $^{\dagger}$ The cooperative WFRFT-DM scheme; $^{\ddagger}$ The independent WFRFT-DM scheme.}\\
\end{table*}

{

$\bullet$ Impact of WFRFT Parameters' Leakage

In this paper, the WFRFT parameters are presumed to be securely shared between Alice and Bobs. In practice, secure key links from Alice to Bobs should be established before DM transmissions, which can be achieved using advanced encryption algorithms, as shown in Fig. 7. Therefore, the security of our proposed power-efficient multi-beam WFRFT-DM schemes is two-fold: the first is the encryption security of WFRFT parameters, and the second is the essential physical layer security of DM transmission.  The WFRFT transform actually acts as a supplement for the physical layer security of DM transmission, which overcomes the low-power-efficiency drawback of the conventional AN-DM schemes. To illustrate the advantage of the proposed WFRFT-DM schemes, here we consider the worst case where the WFRFT parameters and the sequence lengths are entirely leaked to Eves, and these Eves are able to cooperate with each other.

Taking the proposed WFRFT-DM scheme with independent receivers as an example, Eves will try to operate inverse WFRFT to their received signals with these WFRFT parameters. After inverse WFRFT operation, the received signal of the $v$-th Eve ($v=1,2,\cdots,V$) in (\ref{eq_Y_E2}b) can be modified as
\begin{subequations}
\label{eq_Y_E1_discus}
\begin{align}
{{\bf{y}}'}_{v}^{\rm{E}}  &= {\mathscr{F}}^{\alpha_{v}^{\rm{E}}}\left(\sqrt {{P_s}} \sum\limits_{k'=1}^{K}{{\varrho}_{v,k'}}{\mathscr{F}}^{\alpha_{k'}^{\rm{A}}}({\bf{s}}_{k'}) + {\boldsymbol{\xi}}_{v}^{\rm{E}}\right) \\
&= \sqrt {{P_s}} \sum\limits_{k'=1}^{K}{{\varrho}_{v,k'}}{\mathscr{F}}^{\alpha_{v}^{\rm{E}}}{\mathscr{F}}^{\alpha_{k'}^{\rm{A}}}({\bf{s}}_{k'}) + {\mathscr{F}}^{\alpha_{v}^{\rm{E}}}({\boldsymbol{\xi}}_{v}^{\rm{E}})
\end{align}
\end{subequations}
where  ${\alpha_{v}^{\rm{E}}}$ denotes the WFRFT parameter of the $v$-th Eve. Without loss of generality, we further suppose the $v$-th Eve is meant to eavesdrop the $k$-th  Bob's information, which means
\begin{equation}
\label{eq_alpha_E_discus}
\alpha_{v}^{\rm E}=\alpha_{k}^{\rm B}=-\alpha_{k}^{\rm A}
\end{equation}
Then the signal in (\ref{eq_Y_E1_discus}) can be further written as
\begin{subequations}
\label{eq_Y_E2_discus}
\begin{align}
&{{\bf y}'}_{v,k}^{\rm{E}}  = \sqrt {{P_s}} \sum\limits_{k'=1}^{K}{{\varrho}_{v,k'}}{\mathscr{F}}^{-\alpha_{k}^{\rm{A}}}{\mathscr{F}}^{\alpha_{k'}^{\rm{A}}}({\bf{s}}_{k'}) + {\mathscr{F}}^{\alpha_{v}^{\rm{E}}}({\boldsymbol{\xi}}_{v}^{\rm{E}})\\
& = \sqrt {{P_s}} {{\varrho}_{v,k}}{\mathscr{F}}^{-\alpha_{k}^{\rm{A}}}{\mathscr{F}}^{\alpha_{k}^{\rm{A}}}({\bf{s}}_{k}) \\
&~~~+ \sqrt {{P_s}} \sum\limits_{k'\neq k}{{\varrho}_{v,k'}}{\mathscr{F}}^{-\alpha_{k}^{\rm{A}}}{\mathscr{F}}^{\alpha_{k'}^{\rm{A}}}({\bf{s}}_{k'}) +  {\mathscr{F}}^{\alpha_{v}^{\rm{E}}}({\boldsymbol{\xi}}_{v}^{\rm{E}})\\
&= \underbrace{\sqrt {{P_s}} {{\varrho}_{v,k}} {\bf{s}}_{k}}_{{\rm{Distorted~Signal}}} 
+ \underbrace{\sqrt {{P_s}} \sum\limits_{k'\neq k}{{\varrho}_{v,k'}} {\mathscr{F}}^{\alpha_{k'}^{\rm{A}}-\alpha_{k}^{\rm{A}}}( {\bf{s}}_{k'})}_{{\rm{Equivalent~Mixed~Noise}}} 
+ \underbrace{ {\boldsymbol{\xi}'}_{v}^{\rm{E}}}_{{\rm{AWGN}}}
\end{align}
\end{subequations}

Observing (\ref{eq_Y_E2_discus}d), there are two cases to discuss:

\begin{enumerate}
\item The first is that the $v$-th Eve's location is the same as the $k$-th Bob's, which, as a matter of fact, is highly unlikely in practice. In this case, ${\bf{h} }({R_v^{\rm{E}}},\theta _v^{\rm{E}}) = {\bf{h} }(R_k^{\rm{B}},\theta _k^{\rm{B}})$,  ${{\varrho}_{v,k}}= 1$ and 
${{\varrho}_{v,k'}}= 0 $ ($ k'\neq k$), which can be obtained from (\ref{eq_P_H_relation}) and (\ref{eq_Rho_v}). Just as the conventional AN-DM schemes, the transmission security will fail in this case. However, it is worth noting that the WFRFT parameters are assumed to be fixed values for convenience, but are  actually compatible to the dynamic updating strategies of WFRFT parameters like \cite{Ding_Alterable_WFRFT}\cite{Cheng_SM_WFRFT}. Therefore, it is still possible to achieve transmission security via dynamic updating WFRFT parameters when the $v$-th Eve's location is the same as the $k$-th Bob's.  By contrast, the transmission security  of the conventional AN-DM schemes will definitely fail when the $v$-th Eve's location is the same as the $k$-th Bob's.

\item The second is a much more practical case where the $v$-th Eve's location is away from the $k$-th Bob's, which was assumed in nearly all of the previous DM-related literature \cite{Ding_Review_DM}-\cite{Hafez_SM_DM}. In this case, ${\bf{h} }({R_v^{\rm{E}}},\theta _v^{\rm{E}})\neq {\bf{h} }(R_k^{\rm{B}},\theta _k^{\rm{B}})$, ${{\varrho}_{v,k}}\neq 1$, and  ${{\varrho}_{v,k'}}\neq 0 $ ($ k'\neq k$). Owing to  $\alpha_{k'}^{\rm{A}}\neq \alpha_{k}^{\rm{A}}$ (${k'}\neq k$), the received signal of the $v$-th Eve in (\ref{eq_Y_E2_discus}d) is distorted by the coefficient $\varrho _{v,k}$ and interfered by other users' signals. Therefore, even though Eves acquire the knowledge of WFRFT parameters and the sequence lengths, they cannot recover the confidential information as long as Eves' locations are not the same as Bobs'.

\end{enumerate}

Overall, considering the improved power efficiency and the above-discussed cases, our proposed WFRFT-DM schemes still outperform the conventional AN-DM schemes even with leakage of WFRFT parameters.

$\bullet$ Implementing Complexity

}

\begin{figure*}
\centering
\includegraphics[angle=0,width=0.9\textwidth]{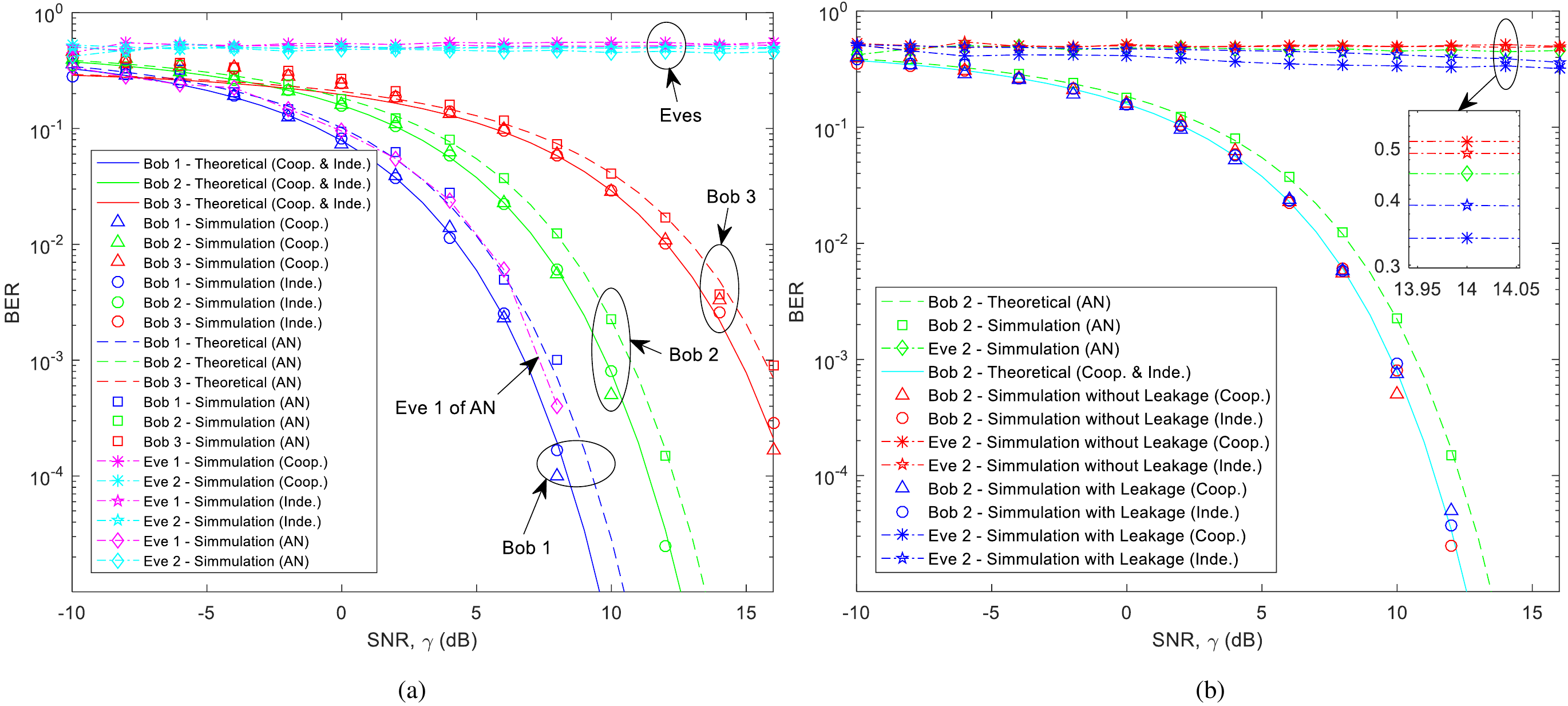}
\caption{{BER performances versus SNR (dB) for the proposed multi-beam WFRFT-DM schemes and the conventional AN-DM scheme. (a) Without leakage of WFRFT parameters; (b) With leakage of WFRFT parameters.}}
\end{figure*}

\begin{figure*}
\centering
\includegraphics[angle=0,width=0.9\textwidth]{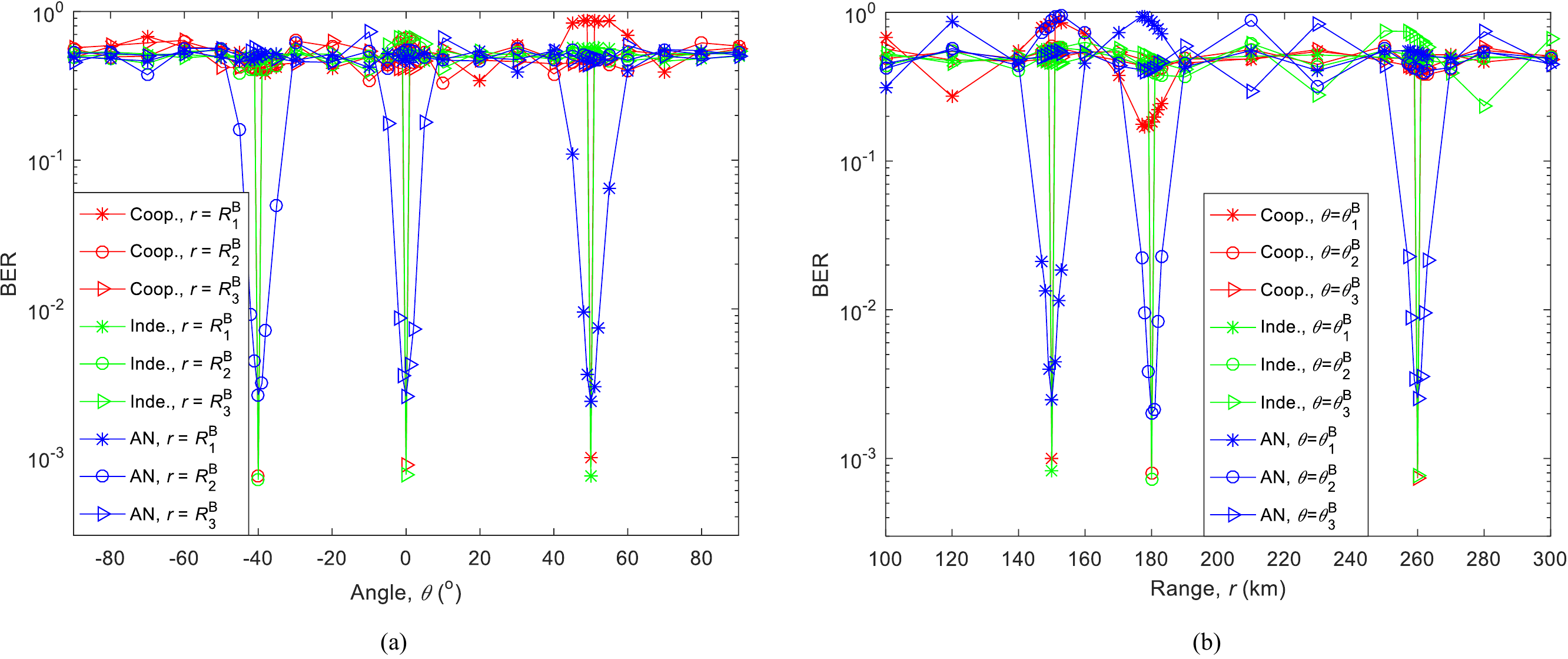}
\caption{{BER performances versus (a) angle and (b) range for the proposed multi-beam WFRFT-DM schemes and the conventional multi-beam AN-DM scheme.}}
\end{figure*}

The proposed multi-beam WFRFT-DM schemes can improve the power efficiency by avoiding the artificial noise. Nevertheless, some additional WFRFT operations have to be performed, even though they do not require calculating the orthogonal matrix, which is necessary for the AN-based method. As for the calculation of WFRFT, from (\ref{eq_DFT}), the WFRFT is based on fundamental DFT, which can be calculated digitally using fast Fourier transform algorithm \cite{Mei_Computation_WFRFT}. On the other hand, since the inverse WFRFT can be easily achieved by substituting the WFRFT parameter $\alpha$ with $-\alpha$, the architectures of calculating WFRFT  are the same for both Alice and Bobs, which can also reduce the implementing complexity to some extent.

\section{Simulation Results}

In this section, we will simulate the BER, secrecy rate, and robustness of the proposed multi-beam WFRFT-DM schemes. The detailed simulation parameters are listed in Table II.

\begin{figure*}
\centering
\includegraphics[angle=0,width=1\textwidth]{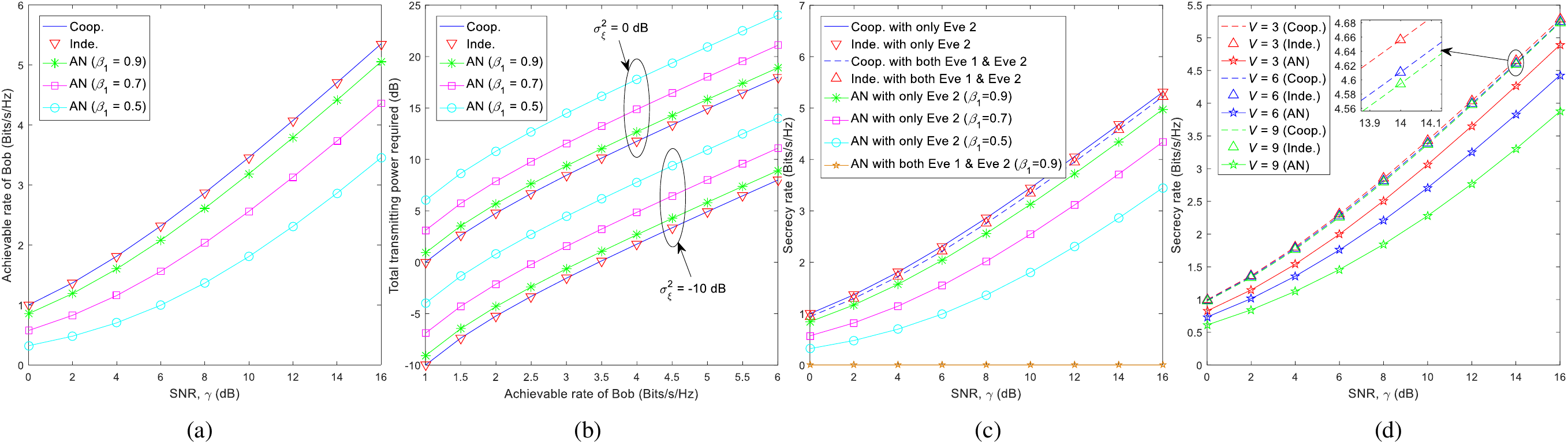}
\caption{{ Secrecy rate performances of the proposed multi-beam WFRFT-DM schemes and the conventional AN-DM scheme. (a) Achievable rate of Bob versus SNR; (b) Total transmitting power required to achieve a given rate of Bob; (c) Secrecy rate versus SNR ($V<K$); (d) Secrecy rate versus SNR ($V\ge K$).}}
\end{figure*}

\subsection{BER}
{Fig. 8(a) shows the BER performances of Bobs and Eves versus SNR (dB) without leakage of WFRFT parameters,} where the baseband modulation modes of Bob 1, Bob 2, and Bob 3 are set as BPSK, QPSK, and 8PSK, respectively. The power splitting factor of the AN-DM scheme is set as $\beta_1 = 0.9$. Several observations can be generalized from {Fig. 8(a)}: 1) In order to achieve the same BER level (e.g., when BER $=10^{-3}$), the SNR required for the proposed WFRFT-DM schemes is approximately $1$ dB less than that of the conventional AN-DM scheme; 2) The BER of Eve 1 is as good as that of Bob 1 for the conventional AN-DM scheme, which is unable to tackle the neighbor eavesdropper; by contrast, the proposed schemes are capable of preventing the neighbor eavesdropper; 3) Different baseband modulations can be applied into different Bobs, which means the proposed schemes are also capable of independent transmissions.

{Fig. 8(b) describes the BER comparisons without/with WFRFT parameters' leakage. For clarity, only the BER curves of Bob 2 and Eve 2 are depicted. On the one hand,  even if the WFRFT parameters are leaked to Eves, the BER of Bob is still as good as the case without leakage. On the other hand, when the WFRFT parameters are leaked to Eve, the BER of Eve is only slightly better than the case without leakage but not better enough to threat the secure transmission between Alice and Bob. Therefore, the transmission security can be also guaranteed in the worst case, which verifies the analysis in Section V.E.}

{Fig. 9(a) and (b)} illustrate the BER performances versus angle and range for the proposed WFRFT-DM schemes and the conventional AN-DM scheme, respectively. In this case, the baseband modulations of three Bobs are set as QPSK, the SNR is $\gamma=10$ dB, and the power splitting factor of the AN-DM scheme is set as $\beta_1 = 0.9$.

From {Fig. 9}, on the one hand, the BER lobes of the proposed schemes are much more slim than that of the conventional AN-DM scheme in both angle and range dimensions. This phenomenon illustrates that the security of the proposed schemes can also hold even if the Eves are located very close to Bobs, while the security of the conventional AN-DM scheme is weakened with the neighbor eavesdropper. On the other hand, the BER performances of Bobs of the proposed schemes are better than that of the conventional AN-DM scheme, which is because the inserted AN consumes a part of the total transmitting power for the AN-based scheme.

\subsection{Secrecy Rate}
{Fig. 10(a)} illustrates the achievable rate of Bob versus SNR (dB) for the proposed multi-beam WFRFT-DM schemes and the conventional AN-DM scheme. As expected, the achievable rates of Bob for the proposed multi-beam WFRFT-DM schemes are much higher than that of the conventional AN-DM scheme, especially when the power splitting factor $\beta_1$ of the AN-DM scheme is smaller.

{Fig. 10(b)} depicts the total transmitting power required for the proposed schemes and the conventional AN-DM scheme in order to achieve a given rate for Bob with fixed AWGN variance. It is observed that the total transmitting power required for the proposed scheme is much smaller than that of the conventional AN-DM scheme, especially with smaller power splitting factor.

It is demonstrated by {Fig. 10(a) and Fig. 10(b)} that the proposed multi-beam WFRFT-DM schemes are much more power-efficient than the traditional multi-beam AN-DM scheme.

Moreover, {Fig. 10(c)} shows the secrecy rate versus SNR (dB) for the proposed multi-beam WFRFT-DM schemes and the conventional AN-based scheme. Whatever value the splitting factor $\beta_1$ takes, the secrecy rates of the proposed schemes are much higher than that of the conventional AN-DM scheme with a fixed normalized total transmitting power. { {Given the specific locations of Bobs and Eves in Table II, the secrecy rates decrease with larger number of Eves.}} Additionally, the secrecy rate of the conventional AN-DM scheme with both Eve 1 and Eve 2 maintains zero whatever the SNR takes, which means no positive secrecy rate will be achieved for the conventional AN-DM scheme when one of the Eves is located at the same place as one of the Bobs. By contrast, the secrecy rates of the proposed schemes are always positive wherever the Eves are located.

{Fig. 10(d) presents the secrecy rate performance when Eves outnumber Bobs. In our simulation, the $V=9$ locations of Eves are randomly generated, which are $(259~{\rm{km}},107^\circ)$, $(221~{\rm{km}},-106^\circ)$, $(298~{\rm{km}},-138^\circ)$, $(157~{\rm{km}},41^\circ)$, $(159~{\rm{km}},-69^\circ)$, $(182~{\rm{km}},-34^\circ)$, $(247~{\rm{km}},8^\circ)$, $(229~{\rm{km}},1^\circ)$, and $(188~{\rm{km}},10^\circ)$, respectively. This illustrates that the secrecy rates of the proposed WFRFT-DM schemes are higher than that of the conventional AN-DM scheme. More importantly, along with increasing number of Eves (from $V=3$ to $V=9$ in the simulations), both the secrecy rates of the AN-DM and WFRFT-DM schemes decline, but the secrecy rates of the  proposed WFRFT-DM schemes decrease much slower than the conventional AN-DM scheme, which also verifies the advantage of the proposed schemes. }

\begin{figure*}
\centering
\includegraphics[angle=0,width=0.8\textwidth]{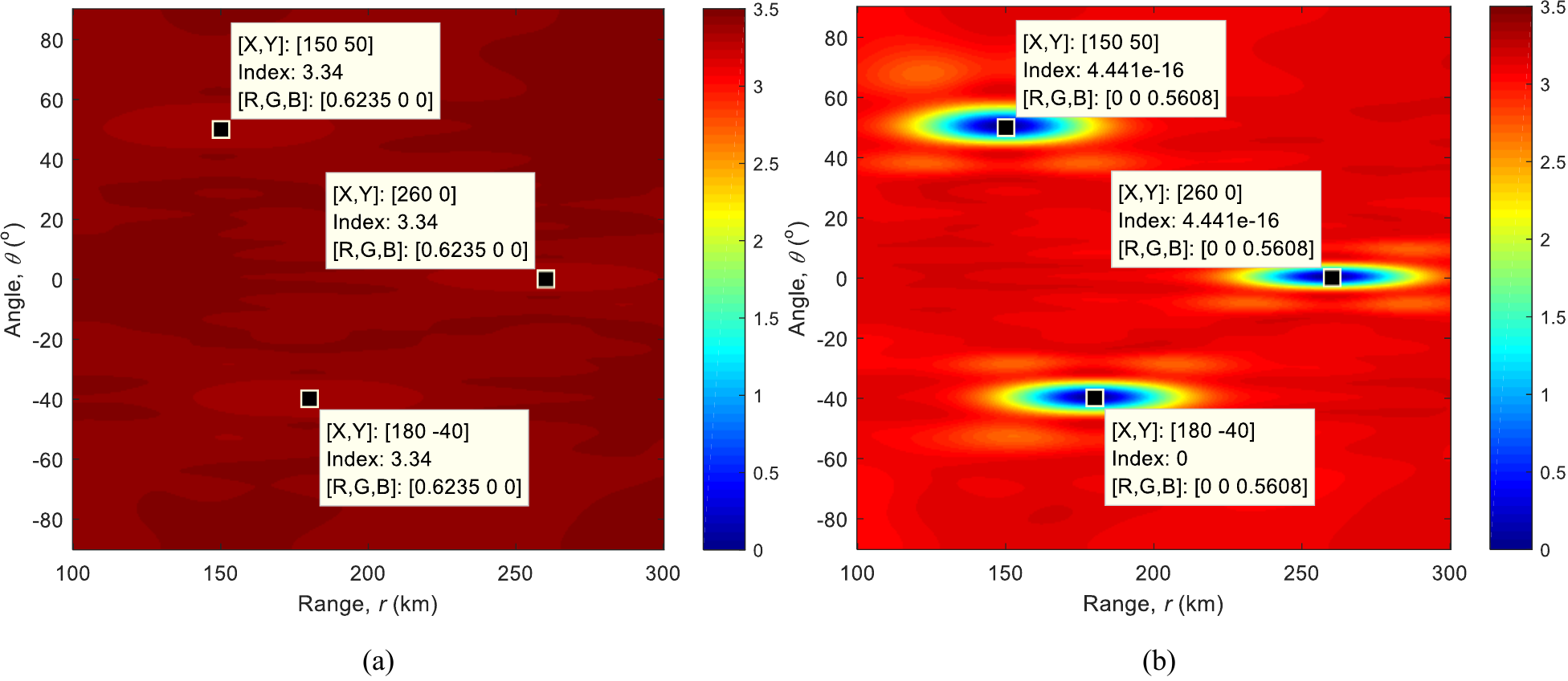}
\caption{{Secrecy rate versus Eve's location. (a) The proposed multi-beam WFRFT-DM schemes; (b) The conventional multi-beam AN-DM scheme.}}
\end{figure*}

\begin{figure*}
\centering
\includegraphics[angle=0,width=0.96\textwidth]{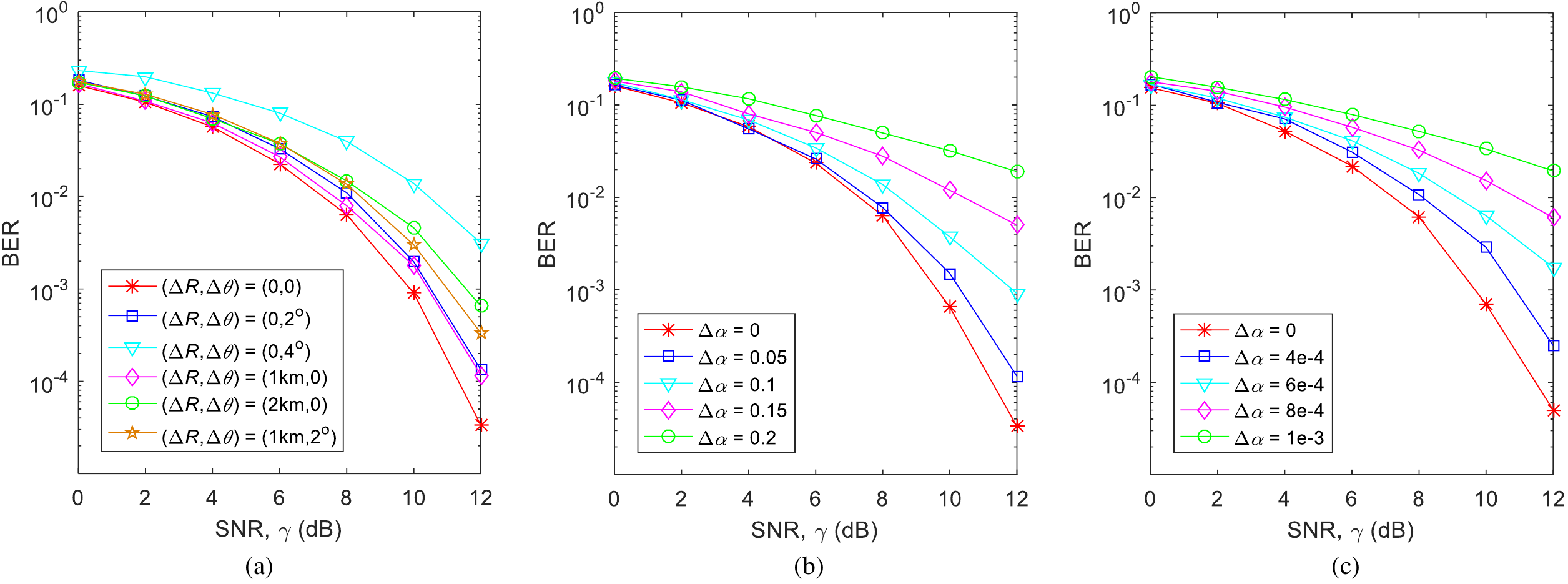}
\caption{Robustness of the proposed multi-beam WFRFT-DM scheme. (a) BER versus SNR (dB) with imperfect estimation of Bobs' locations; (b) BER versus SNR (dB) with mismatched  WFRFT parameter $\Delta\alpha$ ($M_V=N_V=[0~0~0~0]^{\rm{T}}$); (c) BER  versus SNR (dB) with mismatched  WFRFT parameter $\Delta\alpha$ ($M_V,N_V\neq[0~0~0~0]^{\rm{T}}$).}
\end{figure*}

In order to investigate the impacts on the secrecy rate when the Eve's location varies, we assume there is only one Eve who may locate anywhere in free space. The secrecy rates versus Eve's location for the proposed WFRFT-DM schemes and the conventional AN-DM scheme are shown in {Fig. 11(a) and (b)}, respectively, where $\gamma=10$ dB and $\beta_1=0.9$. From {Fig. 11(a)}, it is observed that a high secrecy rate can be always achieved for the proposed WFRFT-DM schemes wherever the Eve is located. By contrast, the secrecy rate of the AN-DM scheme becomes much lower when the Eve's location is close to any one of Bobs' as shown in {Fig. 11(b)}. Especially, the secrecy rate of the AN-DM scheme will decrease to zero when the Eve's location is the same as any one of Bobs'.

\subsection{Robustness}

The BER performances versus SNR (dB) with imperfect estimation of Bobs' locations, mismatched  WFRFT parameter $\Delta\alpha$ ($M_V=N_V=[0~0~0~0]^{\rm{T}}$), and mismatched WFRFT parameter $\Delta\alpha$ ($M_V,N_V\neq[0~0~0~0]^{\rm{T}}$) are depicted in {Fig. 12(a)-(c)}, respectively. For clarity, only the BER curves of Bob 2 with QPSK modulation are presented, from which the same conclusions can also hold  for Bob 1 and Bob 3.

It can be seen from {Fig. 12(a)} that given a fixed BER performance (e.g., BER $=10^{-3}$), there is only about $0.5$ dB loss of SNR when the estimated angle error is $\Delta\theta=2^{\circ}$ and the estimated range error is ${\Delta R }= 0$ compared with the ideal case with no errors. On the other hand, when the estimated angle error is $\Delta\theta=0$ and the estimated range error is ${\Delta R }= 1~{\rm{km}}$,  there is only about $0.5$ dB loss of SNR to achieve a given BER performance (e.g., BER $=10^{-3}$) compared with the ideal case with no errors. When there are both angle and range errors, such as $({\Delta R},\Delta\theta)=(1~{\rm{km}},2^{\circ})$, only $1$ dB loss of SNR occurs compared with the ideal case. Therefore, as long as the estimated angle and range errors are less than $({\Delta R},\Delta\theta)=(1~{\rm{km}},2^{\circ})$, at most $1$ dB additional SNR is required to achieve the same BER as the ideal case.

From {Fig. 12(b) and Fig. 12(c)}, two observations can be obtained. First, as expected, the BER performance of the proposed WFRFT-DM schemes will worsen along with larger mismatched WFRFT parameter $\Delta\alpha$. Second, in order to achieve a given BER performance (eg., BER $=10^{-3}$), there is only $0.5$ dB SNR loss with $\Delta\alpha=0.05$ when the single-parameter WFRFT is adopted ($M_V=N_V=[0~0~0~0]^{\rm{T}}$); by contrast, almost $1$ dB SNR loss will occur even with smaller $\Delta\alpha=4\times10^{-4}$ when the multi-parameter WFRFT is adopted ($M_V,N_V\neq[0~0~0~0]^{\rm{T}}$). This means that the proposed DM scheme with single-parameter WFRFT is less sensitive, more robust than using multi-parameter WFRFT. Therefore, as long as the key link from Alice to Bobs can be securely guaranteed, the single-parameter WFRFT is a better choice than multi-parameter WFRFT to improve the robustness of the multi-beam DM system.

\section{Conclusion}
With the assistance of the WFRFT technology, two power-efficient multi-beam WFRFT-DM schemes with cooperative/independent receivers were proposed, respectively. The BER, secrecy rate, and robustness were analyzed and simulated, which verified the advantages of power efficiency and neighbor security for the proposed multi-beam WFRFT-DM schemes over the conventional multi-beam AN-DM scheme. One future work is to investigate the capability of the proposed FDA-based multi-beam WFRFT-DM schemes over multi-path fading channels.

\ifCLASSOPTIONcaptionsoff
  \newpage
\fi

\begin{IEEEbiography}[{\includegraphics[width=1in,height=1.25in,clip,keepaspectratio]{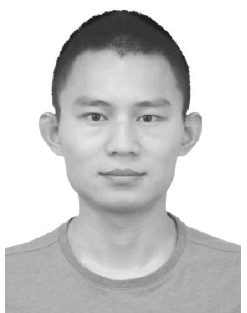}}]{Qian Cheng}
received his B.S. and M.S. degrees in Information and Communication Engineering from Xidian University, Xi'an, China, in 2014, and National University of Defense Technology (NUDT), Changsha, China, in 2016, respectively. 

He is currently pursuing his Ph.D. degree at NUDT; meanwhile, he is a visiting Ph.D. researcher at the Institute of Electronics, Communications and Information Technology (ECIT), Queen's University Belfast, Belfast, U.K. His  research interests include physical layer security and directional modulation.
\end{IEEEbiography}

\begin{IEEEbiography}[{\includegraphics[width=1in,height=1.25in,clip,keepaspectratio]{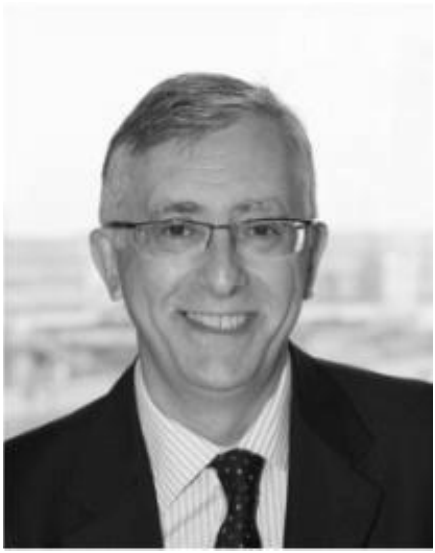}}]{Vincent Fusco}
(S'82-M'82-SM'96-F'04) received the bachelor's degree (Hons.) in electrical and electronic engineering, the Ph.D. degree in microwave electronics, and the D.Sc. degree from Queen's University Belfast (QUB), Belfast, U.K., in 1979, 1982, and 2000, respectively. 

He is currently a Chief Technology Officer with the Institute of Electronics, Communications and Information Technology (ECIT), QUB. He has authored more than 450 scientific papers in major journals and in referred international conferences and two textbooks. He holds patents related to self-tracking antennas and has contributed invited papers and book chapters. His current research interests include advanced front-end architectures with enhanced functionality, active antenna, and front-end MMIC techniques. 

Dr. Fusco is a Fellow of the Institution of Engineering and Technology, the Royal Academy of Engineers, and the Royal Irish Academy. He was a recipient of the IET Senior Achievement Award and the Mountbatten Medal, in 2012. He serves on the Technical Program Committee of various international conferences, including the European Microwave Conference.
\end{IEEEbiography}

\begin{IEEEbiography}[{\includegraphics[width=1in,height=1.25in,clip,keepaspectratio]{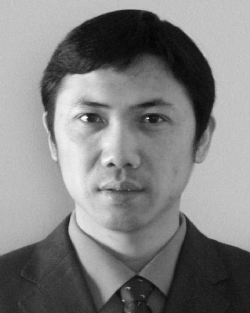}}]{Jiang Zhu}
received the B.S., M.S., and Ph.D. degrees in electrical engineering from the National University of Defense Technology (NUDT), Changsha, China, in 1994, 1997, and 2000, respectively. From 2000 to 2004, he was a lecturer in communication engineering at the NUDT. He was a visiting scholar at the University of Calgary, AB, Canada from April 2004 to July 2005. From 2005 to 2008, he was an associate professor in communication engineering at the NUDT. 

Since 2008, he has been with the NUDT as a full professor in the School of Electronic Science and Engineering. His current research interests include wireless high speed communication technology, satellite communication, physical layer security and wireless sensor network.
\end{IEEEbiography}

\begin{IEEEbiography}[{\includegraphics[width=1in,height=1.25in,clip,keepaspectratio]{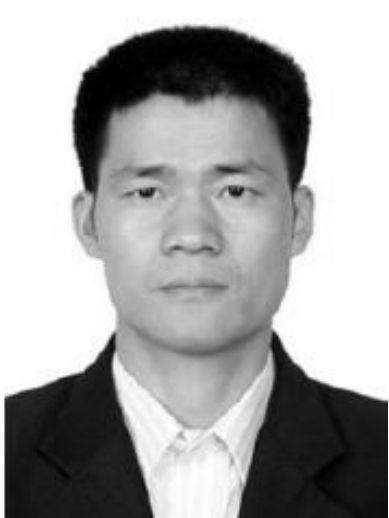}}]{Shilian Wang}
received his B.S. and Ph.D. degrees in information and communication engineering from National University of Defense Technology (NUDT), Changsha, China, in 1998 and 2004, respectively. Since 2004, he continued research in wireless communications at NUDT, where he later became a Professor. From 2008 to 2009, he was a visiting scholar with the Department of Electronic and Electrical Engineering at Columbia University (CU), New York, USA.

He is currently the Head of the Laboratory of Advanced Communication Technology at the School of Electronic Science, NUDT. He has authored or co-authored two books, 26 journal papers, and 20 conference papers. His research interests include wireless communications and signal processing theory, including chaotic spread spectrum and LPI communications, physical layer security, spatial modulation,  deep learning and its applications in communication sensing, etc.
\end{IEEEbiography}

\begin{IEEEbiography}[{\includegraphics[width=1in,height=1.25in,clip,keepaspectratio]{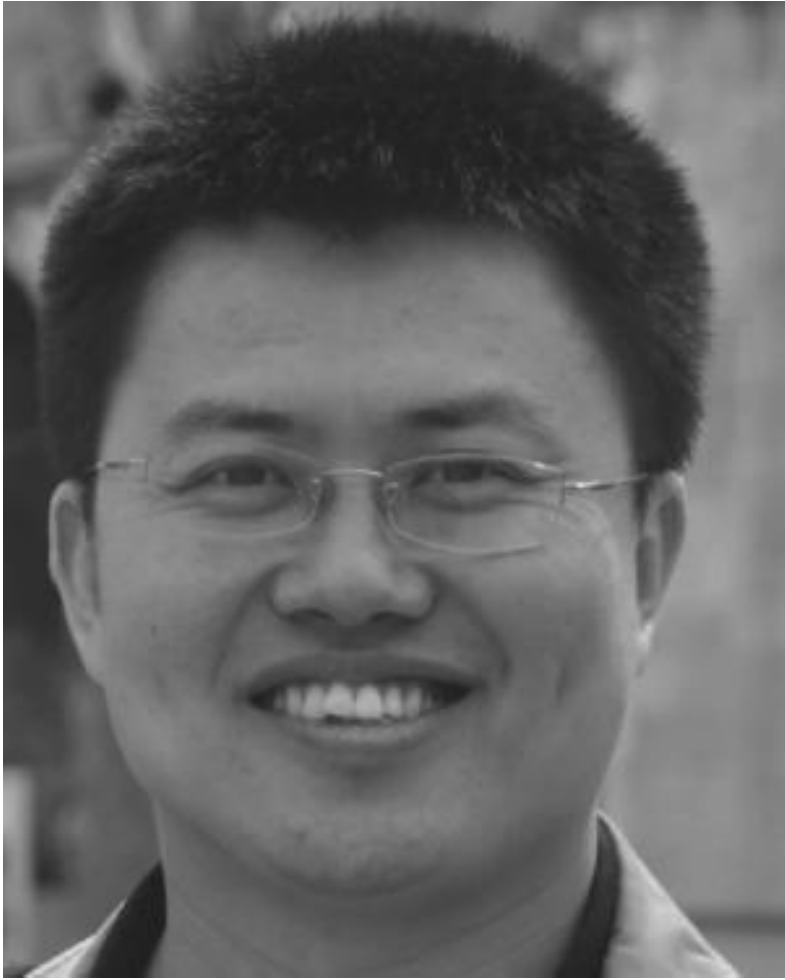}}]{Fanggang Wang}
(S'10-M'11-SM'16) received the B.Eng. and Ph.D. degrees from the School of Information and Communication Engineering, Beijing University of Posts and Telecommunications, Beijing, China, in 2005 and 2010, respectively. He was a Post-Doctoral Fellow with the Institute of Network Coding, The Chinese University of Hong Kong, Hong Kong, from 2010 to 2012. He was a Visiting Scholar with the Massachusetts Institute of Technology from 2015 to 2016, and with the Singapore University of Technology and Design in 2014. 

He is currently a Professor with the State Key Laboratory of Rail Traffic Control and Safety, School of Electronic and Information Engineering, Beijing Jiaotong University. His research interests are in wireless communications, signal processing, and information theory. He served as an Editor for the IEEE Communications Letters and a technical program committee member for several conferences.
\end{IEEEbiography}

\vfill


\begin{thebibliography}{1}


\bibitem{Ding_Review_DM}
Y. Ding and V. Fusco, ``A review of directional modulation technology," \emph{Int. J. Microw. Wirel. Tech.}, vol. 8, no. 7, pp. 981-993, Nov. 2016.

\bibitem{Daly_PA_DM1}
M. P. Daly and J. T. Bernhard, ``Directional modulation technique for phased arrays," \emph{IEEE Trans. Ante. \& Propag.}, vol. 57, no. 9, pp. 2633-2640, Sep. 2009.

\bibitem{Daly_PA_DM2}
M. P. Daly and J. T. Bernhard, ``Demonstration of directional modulation using a phased array," \emph{IEEE Trans. Ante. \& Propag.}, vol. 58, no. 5, pp. 1545-1550, May 2010.

\bibitem{Ding_BER_DM}
Y. Ding and V. Fusco, ``BER-driven synthesis for directional modulation secured wireless communication," \emph{Int. J. Microw. Wirel. Tech.}, vol. 6, no. 2, pp. 139-149, 2014.

\bibitem{Hong_RF}
T. Hong, M.-Z. Song, and Y. Liu, ``RF directional modulation technique using a switched antenna array for physical layer secure communication applications," \emph{Progress In Electr. Research}, vol. 116, pp. 363-379, May 2011.

\bibitem{Hong_Dual_Beam}
T. Hong, M.-Z. Song, and Y. Liu, ``Dual-beam directional modulation technique for physical-layer secure communication," \emph{IEEE Ante. \& Wirel. Propag. Lett.}, vol. 10, pp. 1417-1420, Dec. 2011.


\bibitem{Hong_Convex_Opt}
T. Hong, X.-P. Shi, and X.-S. Liang, ``Synthesis of sparse linear array for directional modulation via convex optimization,'' \emph{IEEE Trans. Ante. \& Propag.}, vol. 66, no. 8, pp. 3959-3972, Aug. 2018.

\bibitem{Wang_FDA_DM}
W.-Q. Wang, ``DM using FDA antenna for secure transmission," \emph{IET Microw., Ante. \& Propag.}, vol. 11, no. 3, pp. 336-345, Apr. 2017.

\bibitem{Xiong_FDA_DM}
J. Xiong, S. Y. Nusenu, and W.-Q. Wang, ``Directional modulation using frequency diverse array for secure communications," \emph{Wirel. Personal Commun.}, vol. 95, no. 3, pp. 2679-2689, Aug. 2017.

\bibitem{Cheng_Time_FDA_DM}
Q. Cheng, J. Zhu, T. Xie, J. Luo, and Z. Xu, ``Time-invariant angle-range dependent directional modulation based on time-modulated frequency diverse arrays," \emph{IEEE Access}, vol. 5, pp. 26279-26290, Dec. 2017.

\bibitem{Ding_Vector_DM}
Y. Ding and V. Fusco, ``A vector approach for the analysis and synthesis of directional modulation transmitters," \emph{IEEE Trans. Ante. \& Propag.}, vol. 62, no.  1, pp.  361-370, Jan. 2014.

\bibitem{Hu_Robust_DM}
J. Hu, F. Shu, and J. Li, ``Robust synthesis method for secure directional modulation with imperfect direction angle," \emph{IEEE Commun. Lett.}, vol. 20, no. 6, pp. 1084-1087, Jun. 2016.

\bibitem{Hu_Random_FDA_DM}
J. Hu, S. Yan,  F. Shu, J. Wang, J. Li, and Y. Zhang, ``Artificial-noise-aided secure transmission with directional modulation based on random frequency diverse arrays," \emph{IEEE Access}, vol. 5, no. 99, pp. 1658-1667, Jan. 2017.

\bibitem{Ji_Fading_FDA_DM}
S. Ji, W. Wang, H. Chen, and Z. Zheng, ``Secrecy capacity analysis of AN-aided FDA communication over Nakagami-\emph{m} fading," \emph{IEEE Wirel. Commun. Lett.}, vol. 7, no. 6, pp. 1034-1037, Dec. 2018.

{
\bibitem{Qiu_AN_FDA_DM}
B. Qiu, J. Xie, L. Wang, and Y. Wang, ``Artificial-noise-aided secure transmission for proximal legitimate user and eavesdropper based on frequency diverse arrays,'' \emph{IEEE Access}, vol. 6, pp. 52531-52543, 2018.
}

\bibitem{Ding_Orthogonal_MB}
Y. Ding and V. Fusco, ``Orthogonal vector approach for synthesis of multi-beam directional modulation transmitters," \emph{IEEE Ante. \& Wirel. Propag. Lett.}, vol. 14, pp. 1330-1333, Feb. 2015.



\bibitem{Xie_AN_MB}
T. Xie, J. Zhu, and Y. Li, ``Artificial-noise-aided zero-forcing synthesis approach for secure multi-beam directional modulation," \emph{IEEE Commun. Lett.}, vol. 22, no. 2, pp. 276-279, Feb. 2018.


\bibitem{Shu_AN_MB}
F. Shu, L. Xu, J. Wang, W. Zhu, and X. Zhou, ``Artificial-noise-aided secure multicast precoding for directional modulation systems," \emph{IEEE Trans. Veh. Tech.}, vol. 67, no. 7, pp. 6658-6662, Jul. 2018.

\bibitem{Christopher_AN_MB}
R. M. Christopher and D. K. Borah, ``Iterative convex optimization of multi-beam directional modulation with artificial noise," \emph{IEEE Commun. Lett.}, vol. 22, no. 8, pp. 1712-1715, Aug. 2018.

\bibitem{Shu_Robust_MB1}
F. Shu, X. Wu, J. Li, R. Chen, and B. Vucetic, ``Robust synthesis scheme for secure multi-beam directional modulation in broadcasting systems," \emph{IEEE Access}, vol. 4, pp. 6614-6623, Oct. 2016.

\bibitem{Shu_Robust_MB2}
F. Shu, W. Zhu, X. Zhou, J. Li, and J. Lu, ``Robust secure transmission of using main-lobe-integration-based leakage beamforming in directional modulation MU-MIMO systems,'' \emph{IEEE Syst. J.}, vol. 12, no. 4, pp. 3775-3785, Dec. 2018.

\bibitem{Cheng_SVD_DM}
Q. Cheng, V. Fusco, J. Zhu, S. Wang, and C. Gu, ``SVD-aided multi-beam directional modulation scheme based on frequency diverse array," { \emph{arXiv}:1907.11972v1, Jul. 2019.}

\bibitem{Xie_OFDA_MB_DM}
T. Xie, J. Zhu, Q. Cheng, and Y. Guan, ``Secure point-to-multipoint communication using the spread spectrum assisted orthogonal frequency diverse array in free space,''  
\emph{IEICE Trans. Commun.}, { vol. E102.B, no. 6, pp. 1188-1197, Jun. 2019.}


\bibitem{Qiu_MB_FDA_DM}
B. Qiu, M. Tao, L. Wang, J. Xie, and Y. Wang, ``Multi-beam directional modulation synthesis scheme based on frequency diverse array,'' \emph{IEEE Trans. Info. Foren. \& Sec.}, { vol. 14, no. 10, pp. 2593-2606, Oct. 2019.}


\bibitem{Ding_RDA_DM}
Y. Ding and V. Fusco, ``A synthesis-free directional modulation transmitter
using retrodirective array,'' \emph{IEEE J. Sel. Topics Sig. Processing}, vol. 11, no. 2, pp. 428-441, Mar. 2017.





\bibitem{Hafez_Multipath_DM2}
M. Hafez, T. Khattab, T. Elfouly, and H. Arslan, ``Secure multiple-users transmission using multi-path directional modulation,'' in \emph{Proc. IEEE Int. Conf. Commun. (ICC)}, Kuala Lumpur, Malaysia, May 2016, pp. 1-5.

\bibitem{Hafez_SM_DM}
M. Hafez, M. Yusuf, T. Khattab, T. Elfouly, and H. Arslan, ``Secure spatial multiple access using directional modulation,'' \emph{IEEE Trans. Wirel. Commun.}, vol. 17, no. 1, pp. 563-573, Jan. 2018.








\bibitem{Mei_Research_on_4WFRFT}
L. Mei, X. Sha, Q. Ran, and N. Zhang, ``Research on the application of 4-weighted fractional Fourier transform in communication system,'' \emph{Sci. China: Info. Sci.}, vol. 53, no. 6, pp. 1251-1260, Jun. 2010.

\bibitem{Fang_Guaranteeing_PLS_WFRFT}
X. Fang, X. Sha, and L. Mei, ``Guaranteeing wireless communication secrecy via a WFRFT-based cooperative system,'' \emph{China Commun.}, vol. 12, no. 9, pp. 76-82, Sep. 2015.

\bibitem{Fang_Secret_Commu_WFRFT}
X. Fang, X. Sha, and Y. Li, ``Secret communication using parallel combinatory spreading WFRFT,'' \emph{IEEE Commun. Let.}, vol. 19, no. 1, pp. 62-65, Jan. 2015.

\bibitem{Luo_Polarization_WFRFT1}
Z. Luo, H. Wang, K. Zhou, and W. Lv, ``Combined constellation rotation with weighted FRFT for secure transmission in polarization modulation based dual-polarized satellite communications,'' \emph{IEEE Access}, vol. 5, pp. 27061-27073, Nov. 2017.

\bibitem{Luo_Polarization_WFRFT2}
Z. Luo, H. Wang, and K. Zhou, ``Physical layer security scheme based on polarization modulation and WFRFT processing for dual-polarized satellite systems,'' \emph{KSII Trans. Internet \& Info. Syst.}, vol. 11, no. 11, pp. 5610-5624, Nov. 2017.

\bibitem{Fang_PLS_Cooperation_WFRFT1}
X. Fang, N. Zhang, X. Sha, D. Chen, X. Wu, and X. S. Shen, ``Physical layer security: a WFRFT-based cooperation approach,'' in \emph{Proc. IEEE Int. Conf. Commun. (ICC)}, Paris, France, May 2017, pp. 1-6.

\bibitem{Fang_PLS_Cooperation_WFRFT2}
X. Fang, N. Zhang, S. Zhang, D. Chen, X. Sha, and X. Shen, ``On physical layer security: weighted fractional Fourier transform based user cooperation,'' \emph{IEEE Trans. Wirel. Commun.}, vol. 16, no. 8, pp. 5498-5510, Aug. 2017.

\bibitem{Ding_Alterable_WFRFT}
B. Ding, L. Mei, and X. Sha, ``Secure communication system based on alterable-parameter 4-weighted fractional Fourier transform,'' \emph{Info. Tech. Journal}, vol. 9, pp. 158-163, 2010.

\bibitem{Liang_Multi_Para_WFRFT}
Y. Liang, X. Da, R. Xu, L. Ni, D. Zhai, and Y. Pan, ``Research on constellation-splitting criterion in multiple parameters WFRFT modulations,'' \emph{IEEE Access}, vol. 6, pp. 34354-34364, Jun. 2018.

\bibitem{Cheng_SM_WFRFT}
Q. Cheng, J. Zhu, and J. Luo, ``Secure spatial modulation based on dynamic multi-parameter WFRFT,'' \emph{IEICE Trans. Commun.}, vol. E101.B, No. 11, pp. 2304-2312, 2018.

\bibitem{Sha_Hybrid_CDMA_WFRFT}
X. Sha, X. Qiu, and L. Mei, ``Hybrid carrier CDMA communication system based on weighted-type fractional Fourier transform,'' \emph{IEEE Commun. Let.}, vol. 16, no. 4, pp. 432-435, Apr. 2012.

\bibitem{Mei_Hybrid_Frequency_WFRFT}
L. Mei, Z. Wang, X. Sha, X. Wang, and N. Zhang, ``BER analysis of hybrid carrier system based on WFRFT with carrier frequency offset,'' \emph{Electro. Let.}, vol. 51, no. 21, pp. 1708-1709, Oct. 2015.

\bibitem{Wang_Power_Allocate_WFRFT}
Z. Wang, L. Mei, X. Wang, X. Sha, and V. C. M. Leung, ``On the performance of hybrid carrier system based on WFRFT with power allocation,'' \emph{IEEE Access}, vol. 6, pp. 29231-29240, May 2018.

\bibitem{Wang_NOMA_WFRFT}
X. Wang, F. Labeau, L. Mei, Z. Wang, and X. Sha, ``Performance of uplink WFRFT-based hybrid carrier systems with non-orthogonal multiple access,'' \emph{IET Commun.}, vol. 12, no. 15, pp. 1891-1899, Sep. 2018.

\bibitem{Shao_MC_FDA}
H. Shao, J. Dai, J. Xiong, H. Chen, and W.-Q. Wang, ``Dot-shaped range-angle beampattern synthesis for frequency diverse array," \emph{IEEE Ante. \& Wirel. Propag. Lett.}, vol. 15, pp. 1703-1706, Feb. 2016.

\bibitem{Nusenu_Review_FDA}
S. Y. Nusenu and A. Basit, ``Frequency diverse array antennas: from their origin to their application in wireless communication systems," \emph{J. Computer Networks \& Commun.}, vol. 2018, no. 1, pp. 1-12, May 2018.


\bibitem{Ding_Metrics_DM}
Y. Ding and V. Fusco, ``Establishing metrics for assessing the performance of directional modulation systems," \emph{IEEE Trans. Ante. \& Propag.}, vol. 62, no. 5, pp. 2745-2755, May 2014.

\bibitem{Proakis_Digital_Commu}
J. G. Proakis and M. Salehi, ``Optimum receivers for AWGN channels,'' in \emph{Digital Commun.}, 5\textsuperscript{th} ed., New York, NY, USA: McGraw-Hill, 2007, pp. 190-195.

\bibitem{Shu_DOA_Estimation}
F. Shu, Y. Qin, T. Liu, L. Gui, Y. Zhang, J. Li, and Z. Han, ``Low-complexity and high-resolution DOA estimation for hybrid analog and digital massive MIMO receive array,'' \emph{IEEE Trans. Commun.}, vol. 66, no. 6, pp. 2487-2501, Jun. 2018.

\bibitem{Coluccia_Range_Estimation}
A. Coluccia and A. Fascista, ``On the hybrid TOA/RSS range estimation in wireless sensor networks,'' \emph{IEEE Trans. Wirel. Commun.}, vol. 17, no. 1, pp. 361-371, Jan. 2018.

\bibitem{Mei_Computation_WFRFT}
L. Mei, Q. Zhang, X. Sha, and N. Zhang, ``Digital computation of the weighted-type fractional Fourier transform,'' \emph{Sci. China: Info. Sci.}, vol. 56, no. 7, pp. 1-12, Jul. 2013.



\end{thebibliography}
\end{document}